\documentclass[a4paper,11pt]{article} 
\usepackage{jcappub} 
\usepackage{lineno}

\usepackage{tikz}
\usetikzlibrary{shapes.geometric, arrows}

\usepackage{color}

\usepackage{url}

\usepackage{graphicx}

\newcommand{\changetext}[1]{\textnormal{#1}}
\newcommand{\changetextfn}[1]{{#1}}

\newcommand{\cmnt}[1]{}

\title{\boldmath Weibel Instability-Driven Seed Magnetic Fields during Reionization}

\author[a,b,c]{Jorie McDermott,}

\emailAdd{mcdermott.286@buckeyemail.osu.edu}

\author[a,b,c]{Manami Roy,}

\author[a,b,c]{and Christopher M. Hirata}

\affiliation[a]{Center for Cosmology and Astroparticle Physics, The Ohio State University, 191 West Woodruff Ave, Columbus, Ohio 43210, USA}
\affiliation[b]{Department of Astronomy, The Ohio State University, 140 West 18th Avenue, Columbus, Ohio 43210, USA}
\affiliation[c]{Department of Physics, The Ohio State University, 191 West Woodruff Ave, Columbus, Ohio 43210, USA}

\abstract{Cosmological reionization was a highly out-of-equilibrium event that affected every parcel of the intergalactic medium, making it a candidate for astrophysical generation of intergalactic magnetic fields. During reionization, the first stars and galaxies ionized the surrounding, largely neutral, medium in ever-expanding envelopes. Photoionization from sources on one side of the front, combined with the quadrupolar angular dependence of the photoionization cross section, leads to an anisotropic electron velocity distribution. We investigate instabilities in these reionization fronts as a mechanism to generate seed magnetic fields. The Weibel instability 
has the potential to create a magnetic field from these anisotropies. We calculate the magnitude of the isotropic and anisotropic distribution within a simulated reionization front. We find that the fractional anisotropy can grow to $6\times 10^{-3}$ toward the middle of the ionization front.
We show that the linear growth timescale of the Weibel instability is fast compared to the crossing time of the ionization front ($\sim 2\times 10^5$ seconds). We briefly speculate on the possible non-linear evolution of the instability and the implications for cosmological magnetogenesis.
}

\begin{document}
\maketitle
\flushbottom

\section{Introduction}
\label{sec:intro}

Magnetic fields are widespread in the universe \citep{Zweibel2013}. They have been observed in everything from the large-scale structure of galaxies \changetext{\citep{Wielebinski2005, P_P_Kronberg_1994, fletcher2011magneticfieldsnearbygalaxies, Beck2012, 1992ApJ...387..528K, Bernet_2008, Geach2023} and filaments in the cosmic web \cite{4cc44b637422445da3c0656996a5364a, Botten2020, Tessa2023}} to the smaller scale of the Sun \citep{Donati_2009} and Earth \citep{earth_mag_field}. Magnetic fields are known to play a crucial role in the formation of stars and planets, as well as in the development and stability of observed phenomena such as accretion disks and jets \citep{1979cmft.book.....P, 1983flma....3.....Z, grasso2001magnetic}
For these observed fields to have formed, a seed field must have been present to be amplified by dynamos and other mechanisms present in these objects. \changetext{Observations of TeV blazars and gamma ray bursts even suggest that most of the volume of the intergalactic medium is magnetized (e.g., \cite{Aharonian_2023, g8rr-g7p4}).\footnote{\changetextfn{The blazar constraints are based on the forward-beamed inverse-Compton gamma rays from the beam of $e^+e^-$ pairs produced as TeV gamma rays interact with the extragalactic background light; the intermediate leptons are deflected in the presence of a magnetic field \cite{2007JETPL..85..473N, 2009ApJ...703.1078D, 2010Sci...328...73N}. This interpretation has been debated due to the possibility of electrostatic plasma instabilities affecting the $e^+e^-$ beam even in the unmagnetized case \cite{2012ApJ...752...22B}. (The electromagnetic or Weibel mode is expected to be unimportant both theoretically \cite{2012ApJ...758..102S} and based on laboratory analogs \cite{2013ApJ...770...54M}.) Later work \cite{2025PNAS..12213365A} suggests that the instabilities do not disrupt the beam (leading only to modest angular broadening \cite{2014ApJ...787...49S}), although the instability can be re-enhanced by the presence of background cosmic rays \cite{2025ApJ...989...37A}. In any case, the gamma ray burst constraints are expected to be less sensitive to these effects since the duration is less than the time for the instability to develop.}} This is particularly interesting since these regions have mostly not experienced shocks from infall into filaments or haloes. Such a field might have an origin much earlier in cosmic history.} However, the formation of this ubiquitous seed field \changetext{pervading most of the volume of the Universe} is still largely a mystery.

A few potential mechanisms for the formation of the seed field have been put forth. In general, there are early (top-down) and late (bottom-up) mechanisms for the generation of these seed fields \citep{Durrer_2013, Zweibel2013}. In top-down mechanisms, the seed field is formed prior to large-scale structure formation and therefore is embedded throughout the universe \cite{Kandus_2011}. Bottom-up mechanisms allow seed fields to be formed on small scales during early structure formation \citep{Kulsrud2008, Fenu2011}. The seed fields then spread from these structures to permeate the universe.

One candidate for the formation of these magnetic fields is the Biermann battery \changetext{\citep[e.g.,][]{1950ZNatA...5...65B,  Gnedin_2000, 2004APh....21...59B, Doi_2011, subramanian2018magneticfieldsuniverse, 2021MNRAS.504.2346A, 2021amff.book.....S}}. This battery is a mechanism by which magnetic fields form from a gradient in both temperature and electron density. As the electrons move down the density and temperature gradients, an electric field can be established. This electric field creates magnetic flux and leads to the development of a magnetic field if the temperature and density gradients are not parallel. The gradients required for this mechanism can occur on both small and large scales, allowing it to be a candidate for both top-down and bottom-up magnetic field formation. While this mechanism works well for generating magnetic fields, the magnitude of the magnetic field that it can form on large scales is very small (${B} \approx 10^{-19}\, \rm{G}$) \cite{Gnedin_2000}. They are small enough that they would not, by themselves, explain the size of the magnetic fields that we see today \citep{Kulsrud2008}, including in the intergalactic medium \citep{Neronov_2010, Aharonian_2023}, where only some of the volume contains wind material that could entrain galactic magnetic fields \cite{2001ApJ...556..619F}. Other battery mechanisms have since been proposed, e.g., those driven by radiation pressure \cite{2003PhRvD..67d3505L, 2005A&A...443..367L, 2015MNRAS.453..345D, 2017MNRAS.472.1649D}, cosmic rays \cite{2011ApJ...729...73M, 2024ApJ...965..111Y}, or charged dust grains \cite{2025ApJ...985...55S}. In this paper, we will instead examine the formation of magnetic fields from instabilities.

Magnetic field forming from instabilities during reionization is an example of a bottom-up scenario. Reionization was a process by which the first stars and galaxies ionized much of the universe during early structure formation. Prior to reionization, the universe was largely neutralized during recombination \cite{Wise_2019, 2026enap....4..433C}. 
Due to the universe being neutral prior to reionization, an electron density gradient was created as the stars and galaxies ionized the surrounding medium in expanding bubble-like waves. The ``ionization fronts'' are thin because the mean free path of ionizing photons in neutral gas is very small on cosmological scales. These sites of rapid ionization are out of thermal equilibrium \cite{1994MNRAS.266..343M, 2019ApJ...874..154D, Zeng_2021}, but nevertheless all of the intergalactic medium experienced an ionization front at least once during cosmic history. This makes reionization an excellent candidate for instability-driven magnetogenesis. 

The specific type of instability that we investigate here is the Weibel instability \cite{1959PhRvL...2...83W}. This instability arises from a quadrupolar anisotropic velocity distribution of electrons in a collisionless, initially unmagnetized plasma. Electrons moving in the same direction magnetically attract, so there is a tendency of the plasma to separate into sheets of current pointed opposite directions even as the overall electron density $n_e$ remains fixed; motions of the electrons perpendicular to the sheets tend to suppress this growth, but the instability occurs with positive growth rate for any non-zero anisotropy \cite{1959PhFl....2..337F}. In this case, the anisotropic velocity distribution is sourced by photoionization itself (see, e.g., Ref.~\cite{2022PNAS..11911713Z} for an example in the laboratory).  Such an anisotropy can also be generated in cosmological shocks \cite{2001ApJ...563L..15G, 2003ApJ...599L..57S}; but our focus here is on ionization fronts, because ionization fronts likely reached much of the IGM much earlier than structure formation shocks.
A further discussion on this instability, including the role of collisions, can be found in Section \ref{sec: wi}.

To more concretely investigate the possibility of a Weibel instability at reionization, we solve for the background anisotropic and isotropic parts of the electron distribution as a reionization front moves through. We then compute the linear growth (or decay) rate for magnetic field perturbations, thus showing the viability
of this magnetic field formation mechanism. The growth rates turn out to be very fast compared to the ionization front passage time, implying that the Weibel instability will reach some type of non-linear saturation while material is still being ionized; while we will briefly speculate on the nature of this saturation, a detailed investigation is left to future work.

The remainder of the paper is organized as follows. We examine the conditions of an ionization front and set up the conditions for the Weibel instability in Section \ref{sec:setup}. In Section \ref{sec: evaluating_Gs} we evaluate and discuss the equation for both the isotropic and anisotropic parts of the electron distribution. Next, in Section \ref{sec: results} we review the output of the numerical calculations. In Section \ref{sec: discussion}, we discuss our results and the future steps needed to study magnetic field generation at reionization.

\section{Setup and Methods}
\label{sec:setup}

To better understand how a reionization front could allow for the formation of magnetic fields, we review the structure and how it affects the neutral fraction of H and He. This leads to a discussion of what the Weibel instability is and how it affects a distribution of free electrons. We then use the Boltzmann Equation to evaluate the equations for the evolution of the anisotropic and isotropic distributions of electrons with the incorporation of an instability.

\subsection{The Structure of an Ionization Front}

An ionization front occurs when a mostly neutral gas is hit by photons with enough energy to ionize the atoms. For a specific model, we consider an ionization front at redshift $z = 7$. This is after the peak of reionization as inferred from the CMB polarization \citep{2020A&A...635A..99P}, but prior to the end of reionization as inferred from Lyman-$\alpha$ studies \citep{2015MNRAS.447.3402B, 2020MNRAS.491.1736K, 2022MNRAS.514...55B, 2024MNRAS.533L..49Z}. The majority of the gas was neutral at this time, so as stars and galaxies began to emit radiation, they ionized the surrounding medium. The structure of these ionization fronts would resemble expanding bubbles. As the front moves through the medium, the structure of the neutral and ionized medium evolves. We simulate this structure using a model that is slightly modified from what is explained in \cite{Zeng_2021}.

\begin{figure}[ht!]
    \centering
    \hskip 0.2in
    \includegraphics[width=4.7in]{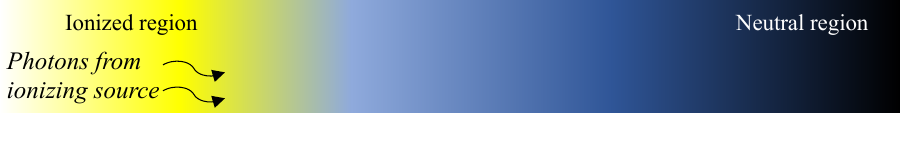}
    \vskip-0.2in
    \centering
    \includegraphics[width=6.1in]{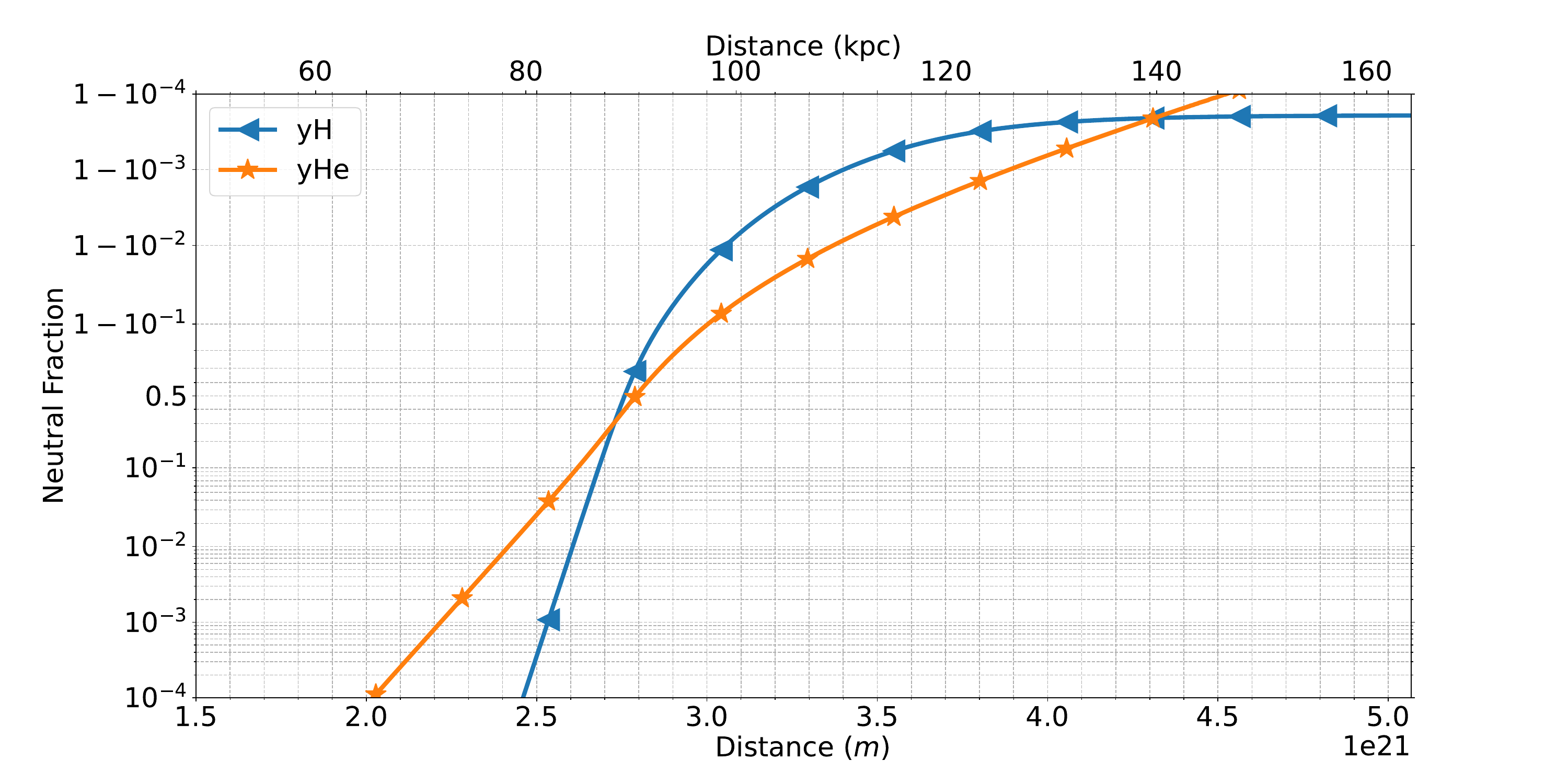}
    \caption{The evolution of the neutral fraction of H and He in the reionization front model. The neutral fraction is plotted on a logit scale on the vertical axis. The horizontal axis is the distance from the ionizing source. As illustrated by the banner, the left side is closest to the ionizing source. Moving right along the horizontal axis is further into the neutral region of the ionization front.}
    \label{fig:neutral_fraction}
\end{figure}

For our reionization front model, we use an ionization velocity of $5 \times 10^6$ m/s and a source temperature of $5 \times 10^4$ K \cite{thermal_evolution_IGM}. To better understand this model, we examine the behavior of the neutral fraction of H and He as the ionization front moves across. The fraction of neutral H and He at different distances from the ionizing source through the model is illustrated in Figure \ref{fig:neutral_fraction}. The ionizing source is located on the left side of this plot, and the right side is the neutral region as described by the schematic diagram at the top of the plot. 

At small distances, we see that the neutral fractions of H and He are both very small because the source has thoroughly ionized the surrounding gas.
The energy required for the first ionization of He is almost double that of what is required for the ionization of H, so the ionization rate is lower for helium ($\Gamma_{\rm HeI}<\Gamma_{\rm HI}$). The neutral fractions would be expected to follow an exponential dependence in the unshielded region, $y_i \propto e^{\Gamma_i z/U}$, where $y_i$ is the neutral fraction of element $i$, $z$ is the distance, and $U$ is the ionization front velocity. Therefore, in this region closer to the source, the neutral fraction of He is greater than the neutral fraction of H. However, the neutral fraction of H increases rapidly and catches up to the neutral fraction of He close to the point where half of the gas is neutral.

After the point where the neutral fraction of both H and He is equal, the He neutral fraction is less than the H neutral fraction for some distance. This is due to the shielding of ionizing photons. At low distances, much of the H is ionized and has absorbed almost all of the photons from the source that are close to the 13.6 eV ionization threshold for H. As a result, only the more energetic ionizing photons can penetrate this region. These have a larger cross-section to photoionize He than H, as the cross-section of helium remains relatively large at higher photon energies, whereas the hydrogen cross-section decreases more steeply. 

At the largest distances in the model, the H and He neutral fraction both increase to nearly 1. This is where none of the ionizing photons have reached, so both H and He are in their pre-reionization, mostly neutral state. There is some ``residual'' ionization $x_e\approx 1.8\times 10^{-4}$ of hydrogen from protons and electrons that were not able to recombine in the cosmological recombination epoch \cite{2000ApJS..128..407S, 2010PhRvD..81h3005G}.

\subsection{Kinetic description of the instability}
\label{sec: wi}



To model the instability quantitatively, we use the linearly perturbed Boltzmann equation, allowing for angular deflection by collisions. The approach follows earlier work on the Weibel instability \cite{1959PhRvL...2...83W, 1959PhFl....2..337F}. Using these constraints, we use the following evolution equation for the distribution function $f$:
\begin{equation}
\frac{\partial{f}}{\partial{t}}(\boldsymbol{r},\boldsymbol{v},t) = -\boldsymbol{v}\cdot\boldsymbol{\nabla_r}{f}-\frac{q_e}{m_e}(\boldsymbol{E}+\boldsymbol{v}\times\boldsymbol{B})\cdot\boldsymbol{\nabla_v}{f}+D_\theta \nabla^2_{ang} f.
\label{eq:dfdt}
\end{equation}
\changetext{(Note that here ${\boldsymbol B}$ is itself a perturbation variable: we do not include a background field.)}

We are using the Weibel instability as our model, so we separate $f(\boldsymbol{r},\boldsymbol{v},t)$ into two components. The first, $f_0$, represents the background distribution and can be separated into isotropic and anisotropic components. The second component is a small perturbation in the distribution function $\delta f$, and has a sinusoidal dependence, $\delta f(\boldsymbol{r},\boldsymbol{v},t) \propto e^{i(k z -\omega t)}$. We relate the components by the following equations:
\begin{equation}
f(\boldsymbol{r},\boldsymbol{v},t) = f_0(\boldsymbol{v}) + \delta f(\boldsymbol{r},\boldsymbol{v},t) ~~~{\rm and}~~~ f_0(\boldsymbol{v})= f_0^{\rm iso}(v) +f_0^{\rm ani}(\boldsymbol{v}).
\label{eq:fcomponents}
\end{equation}

For this work, we assume that the anisotropic part has a quadrupole dependence: there is no net drift of the electrons relative to the center of mass frame, but the velocity dispersion may be larger on some axes (e.g., the $\pm x$-axis) than others (e.g., the $\pm z$-axis). 
For simplicity, we consider a wave vector along a principal axis such as $+\hat{z}$. 
We further assume reflection symmetry of the unperturbed distribution function: $f_0(v_x,v_y,v_z) = f_0(-v_x,v_y,v_z)$. After substituting the components of $f(\boldsymbol{r},\boldsymbol{v},t)$ into Eq.~(\ref{eq:dfdt}) and carrying out the cross-product, we find the following.
\begin{equation}
\left(ik(v_z -u) - D_\theta \nabla_{ang}^2\right) \delta f = \frac{-q_e}{m_e} B_y \left(u \frac{\partial}{\partial v_x} + v_x \frac{\partial}{\partial v_z} - v_z \frac{\partial}{\partial v_x}\right) f_0,
\end{equation}
where $u\equiv (\omega + i\eta)/{k}$ and we have replaced $E_x=u B_y$.
In these equations, $u$ is the phase velocity and $\eta$ is an infinitesimal added to establish the initial conditions for which $\eta \rightarrow 0^+$. We can further simplify this expression by noting that there is no background current, therefore $B_y = \frac{i \mu_0}{k} q_e \int v_x~\delta f d^3 v$. Substituting this into the equation and simplifying, we find that
\begin{equation}
1 = - i \frac{\mu_0 q_e^2}{m_e k} \int d^3\Vec{v}~v_x \left(ik(v_z-u)-D_{\theta}\nabla_{\rm ang}^2\right)^{-1} \left[\left(u \frac{\partial}{\partial v_x} + v_x \frac{\partial}{\partial v_z} - v_z \frac{\partial}{\partial v_x}\right)f_0\right].
\end{equation}
It is helpful to write the right-hand side in terms of a Taylor expansion in $u$. This will allow us to simplify our equation and solve for the isotropic and anisotropic parts of the distribution, which occur under different limits.
We write this equation as $G(u) = k^2/k_{\rm sd}^2$ where the skin depth scale is 
\begin{equation}
k_{\rm sd}=\sqrt{\frac{\mu_0 \bar{n}_e q_e^2}{m_e}},
\end{equation}
and where $G(u)$ is given by
\begin{equation}
G\left(u\right)=-\frac{1}{\bar n_e}\int v_x\left(v_z-u+i\frac{D_\theta}{k} \nabla_{\rm ang}^2\right)^{-1} \left[\left(u\frac{\partial}{\partial v_x} +v_x\frac{\partial}{\partial v_z}-v_z\frac{\partial}{\partial v_x}\right) f_0\right] ~ d^3\Vec{v}.
\label{eq:Gtotal}
\end{equation}
We can then separate $G(u)$ and consider what parts of the distribution in the $\hat{x}$, $\hat{y}$, and $\hat{z}$ directions can be considered part of the background (isotropic) or part of the perturbations in the distribution (anisotropic): 
\begin{equation}
G^{\rm iso}(u) = -\frac{u}{\bar{n}_e} \int_0^{\infty} {dv~v^2 \frac{\partial f_0^{iso}}{\partial v} \int_{S^2} {d\Omega~n_x\left(n_z - \frac{u}{v} + \frac{iD_\theta}{kv} \nabla_{ang}^2\right)^{-1} n_x}}.
\label{eq:Giso}
\end{equation}
This has $G^{\rm iso}(0)=0$, so even when we consider ``small $u$,'' we will take the first-order Taylor expansion in $u$.
For the anisotropic component, we have a contribution even when $u=0$, so we focus on $G^{\rm ani}(0)$: thus the overall expansion for small $u$ or small growth rate is
\begin{equation}
G^{\rm ani}(0) + [G^{\rm iso}{'}(0)]u \approx \frac{k^2}{k_{\rm sd}^2}.
\label{eq:Gk-eq}
\end{equation}
We define 
\begin{equation}
Z \equiv \left(v_x\frac{\partial}{\partial v_z}-v_z \frac{\partial}{\partial v_x}\right) f_0^{ani},
\label{eq:Z}
\end{equation}
so that
\begin{equation}
G^{\rm ani}(0) = -\frac{1}{\bar{n}_e} {\int_0^{\infty}} dv~v^2 \int_{S^2} d\Omega~n_x\left(n_z +\frac{iD_\theta}{kv} \nabla_{ang}^2\right)^{-1} Z.
\label{eq:Gani}
\end{equation}
We can now examine the $G^{\rm iso}$ and $G^{\rm ani}$ functions separately. Solving for each of these will give us a good idea of how the distribution will behave, as well as whether the perturbations will allow magnetic fields to form from anisotropies.

The growth rate of the Weibel instability is thus set by both $G^{\rm iso}{'}(0)$ (which has a positive imaginary part) and thus gives a negative contribution to the left-hand side of Eq.~\ref{eq:Gk-eq} for growing modes) and $G^{\rm ani}(0)$ (which is positive); this is depicted in flow chart form in Figure~\ref{fig:flowchart}.

\renewcommand{\arraystretch}{1.2} 
\tikzstyle{value} = [circle, minimum width=1cm, minimum height=1cm,text centered, draw=white, fill=white]
\tikzstyle{case} = [rectangle, minimum width=2cm, minimum height=1cm, text centered, draw=black, fill=white]
\tikzstyle{arrow} = [thick,->,>=stealth]
\tikzstyle{dasharrow} = [dashed, ->, >=stealth]

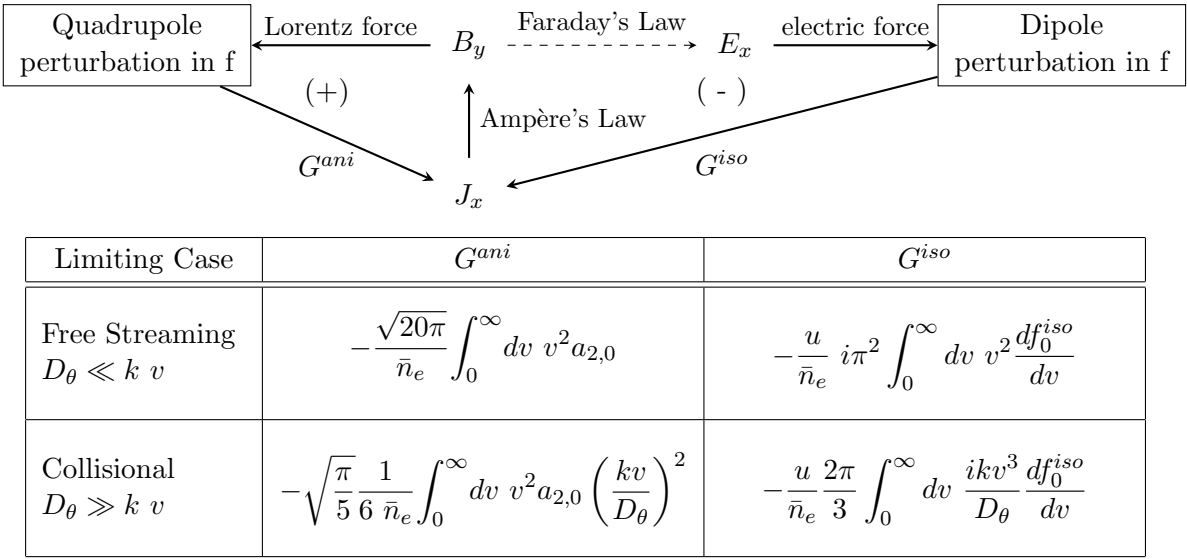
\begin{figure}[ht!]
\centering

\begin{tikzpicture}[node distance=2cm]
\node(B_y)[value]{$B_y$};
\node(E_x)[value, right of=B_y, xshift=1.5cm]{$E_x$};
\node(dipole)[case, right of=E_x, xshift=2.3cm, text width=3cm]{Dipole \\ perturbation in f};
\node(quadrupole)[case, left of=B_y, xshift=-2.5cm, text width=3cm]{Quadrupole perturbation in f};
\node(J_x)[value, below of=B_y]{$J_x$};

\draw[dasharrow](B_y)--node[anchor=south, font=\small]{Faraday's Law}(E_x);
\draw[arrow](B_y)--node[anchor=south, font=\small]{Lorentz force}(quadrupole);
\draw[arrow](J_x)--node[anchor=west, font=\small]{Amp$\grave{\text{e}}$re's Law}(B_y);
\draw[arrow](E_x)--node[anchor=south, font=\small]{electric force}(dipole);
\draw[arrow](quadrupole)--node[anchor=north, yshift=-0.1cm]{$G^{ani}$}node[anchor=south, yshift=0.2cm]{(+)}(J_x);
\draw[arrow](dipole)--node[anchor=north, yshift=-0.1cm]{$G^{iso}$}node[anchor=south, yshift=0.2cm]{( - )}(J_x);

\end{tikzpicture}
\begin{tabular}{|c|c|c|}
\hline
Limiting Case & $G^{ani}$ & $G^{iso}$\\
\hline \hline
\parbox{7em}{Free Streaming \\ $D_\theta \ll k ~ v$} & \parbox{14em}{$$-\frac{\sqrt{20\pi}}{\bar{n}_e}{\int_0^{\infty}} dv~v^2 a_{2,0}$$} & \parbox{14em}{$$-\frac{u}{\bar{n}_e} ~i\pi^2 \int_0^{\infty} {dv~v^2} \frac{d f_0^{iso}}{d v}$$} \\
\hline
\parbox{7em}{Collisional \\ $D_\theta\gg k~v$} & \parbox{14em}{$$-\sqrt{\frac{\pi}{5}}\frac{1}{6 ~ \bar{n}_e}{\int_0^{\infty}} dv~v^2 a_{2,0} \left(\frac{kv}{D_\theta}\right)^2$$} & \parbox{14em}{$$-\frac{u}{\bar{n}_e} \frac{2\pi}{3} \int_0^{\infty} {dv~\frac{ikv^3}{D_\theta} } \frac{d f_0^{iso}}{d v}$$}\\
\hline
\end{tabular}
\label{fig:gani_giso_flowchart_table}
\caption{\label{fig:flowchart}The flowchart for the calculation of $G^{\rm iso}$ and $G^{\rm ani}$. It illustrates the development of both $G^{\rm iso}$ and $G^{\rm ani}$ in terms of the relevant forces and laws, including both positive feedback (``$+$'' loop: current perturbations create a magnetic field that deflects particles to add to the perturbation) and negative feedback (``$-$'' loop: growing currents create a growing magnetic field, which induces an electric field that opposes the current perturbation). The table below the flowchart lists the equations of $G^{\rm ani}$ and $G^{\rm iso}$ in both the free streaming and collisional limits. See Appendix~\ref{sec:calc_sigma} for derivations.}
\end{figure}

\section{Evaluating the Distribution Function}
\label{sec: evaluating_Gs}

This section discusses the considerations of the anisotropic and isotropic distributions. We evaluate the expressions using reionization front data and discuss the results. We start by calculating $G^{\rm iso}$ and discussing the evolution over the distance through the front (Section~\ref{sec:int-Giso}). Next, we evaluate the expression for $G^{\rm ani}$ (Section~\ref{sec:int-Gani}) and the more complex terms it involves --- the source term and the multiple moment (Section~\ref{sec: ani_methods}). We analyze the correlation between these complex terms and the velocities in the front. Lastly, we analyze the relationship of $G^{\rm iso}$ with distance and discuss similarities with the evolution of $G^{\rm iso}$ (Section~\ref{sec: ani_res}).

\subsection{The isotropic contribution, $G^{\rm iso}$}
\label{sec:int-Giso}


Our first goal is to solve for $G^{\rm iso}(u)$, given by Eq.~(\ref{eq:Giso}), in a reionization front. Since this is the isotropic part, it can be computed using a thermal plasma; however, the angular diffusion due to 2-body interactions is significant at the smaller wave numbers, so the anisotropic part will require a numerical solution.

If we only consider the volume integral in Eq.~(\ref{eq:Giso}), we see that in the limit of $u \rightarrow 0$ there are only two pieces: $\sigma$ as described in Eq.~(\ref{eq:sigma_eq}), and a factor of $n_x$. We can then write the integral as
$\int_{S^2} d\Omega ~ n_x \sigma$,
where
\begin{equation}
\sigma(\hat{\boldsymbol n}) = \left(n_z + \frac{iD_\theta}{kv} \nabla_{ang}^2\right)^{-1} n_x
\label{eq:sigma_eq}
\end{equation}
is a function of direction $\hat{\boldsymbol n}$ that depends on the dimensionless combination $D_\theta/kv$.
This can be decomposed using spherical harmonics, $\sigma = \sum_{\ell m} \sigma_{lm}Y_{lm}(\hat{\boldsymbol n})$. Noting that the longitude dependence of Eq.~(\ref{eq:sigma_eq}) is $\propto\cos\phi$, we see that the moments must satisfy
\begin{equation}
\sigma_{\ell,1} = -\sigma_{\ell,-1}
~~~{\rm and}~~~
\sigma_{\ell,m}=0~{\rm for}~|m|\neq 1.
\label{eq:sig-components}
\end{equation}
The hierarchy of $m=1$ moments of $[n_z + (iD_\theta/kv)\nabla_{\rm ang}^2]n_x = \sigma$ can be written as a linear system:
\begin{equation}
    -\sqrt{\frac{2\pi}{3}}\delta_{\ell,1} = \sqrt{\frac{\ell(\ell+2)}{(2\ell+1)(2\ell+3)}}\sigma_{\ell+1,1}+\sqrt{\frac{(\ell-1)(\ell+1)}{(2\ell-1)(2\ell+1)}}\sigma_{\ell-1,1} - \frac{iD_\theta}{kv} \ell (\ell+1) \sigma_{\ell,1},
\label{eq: sigmas_sol}
\end{equation}
which enables us to solve for the $\sigma_{\ell,1}$. For moderate $D_\theta/kv$, the hierarchy can be truncated and solved as a standard linear algebra problem (we use $\ell_{\rm max}= 40$ by default). For small $D_\theta/kv$, very high $\ell$ can contribute and we use the alternative limiting cases described in Appendix~\ref{sec:calc_sigma}.


Using the decomposition that $n_x = 2\sqrt{\pi/6}\,(Y_{1,-1}^\ast - Y_{1,1}^\ast)$, we see that
the integral in Eq.~(\ref{eq:sigma_eq}) will only have contributions from $\ell=1$ and $m=\pm 1$. When we substitute these values of $m$ and $l$ into the integral and simplify in terms of $\sigma_{l,m}$, we find
\begin{equation}
\int_{S^2} d\Omega ~ n_x \sigma = -\frac{2\sqrt{\pi}}{\sqrt{6}} \sigma_{1,1} + \frac{2\sqrt{\pi}}{\sqrt{6}} \sigma_{1,-1} = \frac{-4\sqrt{\pi}}{\sqrt{6}} \sigma_{1,1}.
\label{eq: sigma_11}
\end{equation}
Substituting the solution in Eq.~(\ref{eq: sigma_11}) back into Eq.~(\ref{eq:Giso}), we find
\begin{equation}
G^{\rm iso}\left(u \right) = \frac{u}{\bar{n}_e} \frac{4\sqrt{\pi}}{\sqrt{6}} \int_0^{\infty} {dv~v^2 \sigma_{1,1}} \frac{d f_0^{iso}}{d v}.
\label{eq:Giso-sol}
\end{equation}
Now we can integrate over $v$ and use the result to determine how $G^{\rm iso}$ will act and how it will affect magnetic field generation.
Substituting a Maxwellian distribution for $f_0^{\rm iso}$ into Eq.~(\ref{eq:Giso-sol}) gives
\begin{equation}
G^{\rm iso}\left(u\right) = -\frac{u}{\Bar{n_e}} \frac{4\sqrt{\pi}}{\sqrt{6}} \int_0^{\infty} {dv~v^2 \sigma_{1,1}} \frac{\bar{n_e}v}{(2\pi)^{3/2} \sigma_e^5} e^{{-v^2}/{2\sigma_e^2}},
\label{eq:Giso-lin}
\end{equation}
where $\sigma_e = \sqrt{k_BT_e/m_e}$ is the electron velocity dispersion. Again, this result applies for small $u$: if Taylor-expanded around $u=0$, $G^{\rm iso}$ has no constant term, a linear term given by Eq.~(\ref{eq:Giso-lin}), and higher-order terms that we will not need in this paper.
For numerical computation at small $u$, we can divide by $u$:
\begin{equation}
\frac{G^{\rm iso}\left(u\right)}{u} = - \frac{1}{\pi \sqrt{3}} \int_0^{\infty} {dv} \frac{v^3 ~ \sigma_{1,1}}{\sigma_e^5} e^{{-v^2}/{2\sigma_e^2}}.
\label{eq:Giso/u}
\end{equation}
We then use results from a reionization temperature evolution model \cite{Zeng_2021} as data for use in solving the $G^{\rm iso}/u$ function.

\begin{figure}[ht!]
    \centering
    \includegraphics[width=1\linewidth]{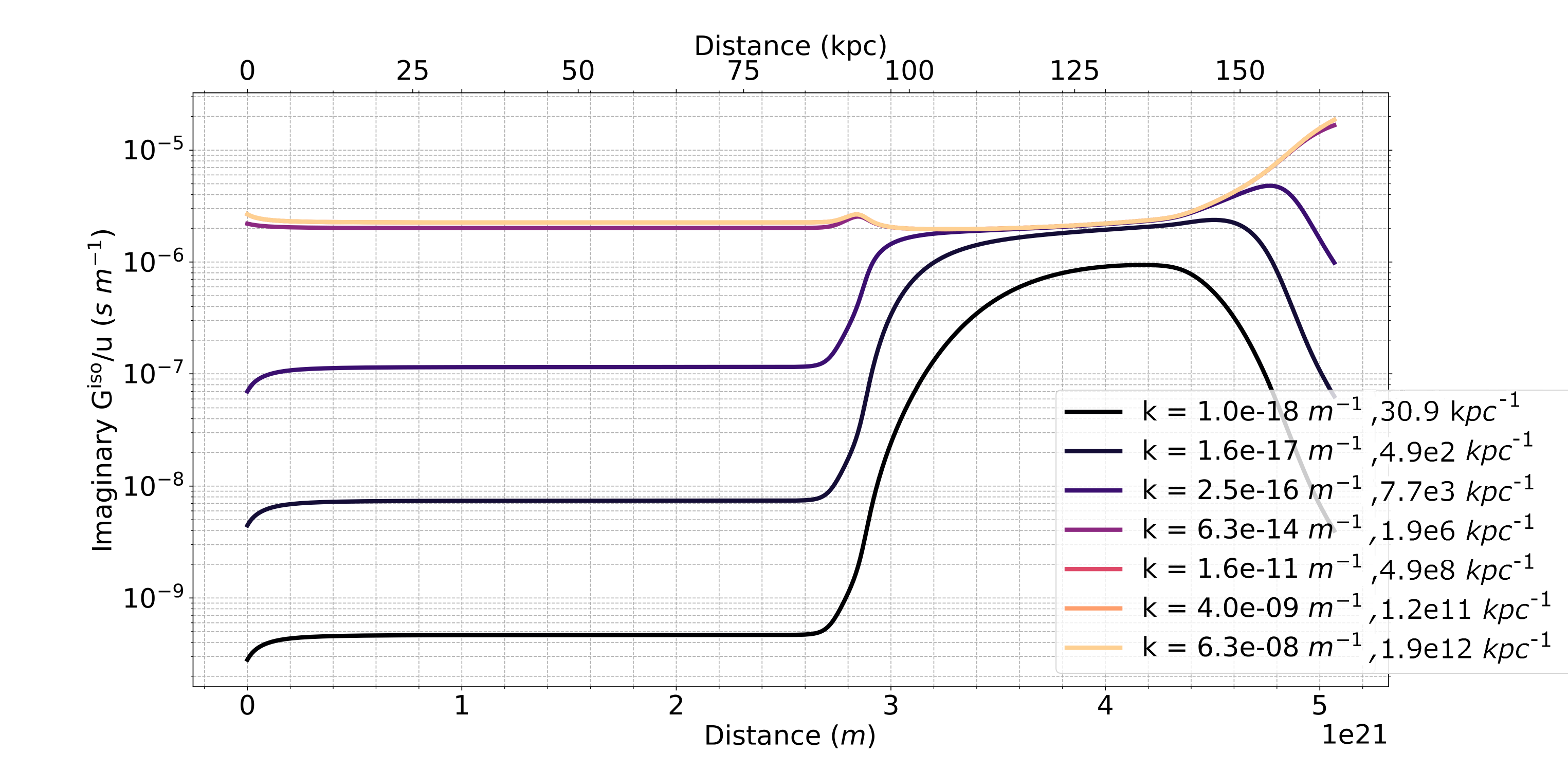}
    \caption{The change in the imaginary $G^{\rm iso}/u$ term across wave number (k) slabs (Eq. \ref{eq:Giso-sol}). The vertical axis is the isotropic distribution value and the horizontal axis is the distance through the reionization front. The increasing k slabs are inversely proportional to different length scales at which we can examine the perturbation. We calculated the sum of $G^{\rm iso}/u$ over 142 velocity bins for this simulation. At the largest distances from the source, the isotropic part of the distribution reaches a maximum for all scales. However, close to the source only small scales (large $k$) are largely isotropic which indicates that the large scales have a significant anisotropic component.}
    \label{fig:giso_k_slabs}
\end{figure}

Note that $\sigma_{1,1}$, and hence $G^{\rm iso}/u$, are positive imaginary. By plotting $G^{\rm iso}/u$ (Figure ~\ref{fig:giso_k_slabs}) in terms of other parameters from the reionization temperature evolution model, we can examine the effects of reionization on $G^{\rm iso}$ and how it changes at different depths through the ionizing front.


We now investigate the behavior of $G^{\rm iso}(u)/u$ as a function of scale (set by the wavenumber $k$) and depth in the ionization front. First, we examine large scales ($k < 6 \times 10^{-14} ~ {\rm m}^{-1}$). As the value of $k$ decreases, $G^{\rm iso}(u)/u$ also decreases in a very steady manner. This is because $G^{\rm iso}(u)/u$ varies between the free-streaming limit (independent of $k$) and the collisional limit ($\propto k$), with a transition when $k^{-1}$ is of order the mean free path. In the neutral region (right side of the figure: distance $\sim 5\times 10^{21}\,$m), the ionization fraction is still mostly determined by residual ionization, but the energy injection from the few ionizing photons that make it to this depth is significant and results in a rapidly increasing electron temperature. One sees the $\propto k$ dependence, with $G^{\rm iso}(u)/u$ increasing sharply as one approaches the front (toward the left) as the temperature rises. However, when we move through the ionization front to a distance of roughly $(3-4)\times10^{21} ~{\rm m}$, $G^{\rm iso}(u)/u$ approaches a plateau at roughly $10^{-6} ~{\rm s} ~ {\rm m}^{-1}$ for all $k\gtrsim 10^{-17}\,{\rm m}^{-1}$. In this region, the mean free path has been reduced by the increasing ionization fraction. Near the ionizing source (distance $<2.5\times 10^{21}\,$m), $G^{\rm iso}(u)/u$ is independent of depth (since the local conditions only depend on the post-front conditions) and is $\propto k$ as expected from the collisional limit, since this region has the shortest electron mean free path.

Next, we can consider the large wavenumbers, $k \geq 6 \times 10^{-14} ~ {\rm m}^{-1}$. These all follow roughly the same distribution and define the maximum value for $G^{\rm iso}/u$ throughout the ionization front. The values for $\sigma_{1,1}$ are within the free-streaming limit as discussed in Appendix~\ref{sec:calc_sigma}, where $G^{\rm iso}(u)/u$ converges to a finite value independent of $k$, which depends principally on the electron temperature. Starting from the neutral side (right-hand side of the plot), we see that $G^{\rm iso}(u)/u$ drops as the first ionizing photons reach the material and heat the electrons. There is then a plateau in the $(3-4)\times 10^{21}\,{\rm m}$ region as the electron temperature is set by the energy injected per ionizing photon. When the ionization fraction reaches a few percent (distance $2.9\times 10^{21}\,{\rm m}$), $G^{\rm iso}(u)/u$ increases slightly before dropping. This is due to the inverse proportionality to temperature of ${G^{\rm iso}(u)}/{u}$. The drop in the temperature in this region as the electrons equilibrate with the ions and neutrals \cite{Zeng_2021} corresponds to the peak in $G^{\rm iso}(u)/u$. Similar to the smaller wavenumbers, the value of $G^{\rm iso}(u)/u$ at large wavenumbers ($k \geq 6 \times 10^{-14} ~ {\rm m}^{-1}$) remains steady close to the ionizing source because it depends only on the post-front conditions.

\subsection{The anisotropic contribution, $G^{\rm ani}$}
\label{sec:int-Gani}

As with the initial equation for $G^{\rm iso}$, $G^{\rm ani}$ in Eq.~(\ref{eq:Gani}) is not currently in a very useful format for numerical evaluation.
%
To start simplifying the equation for $G^{ani}$, we first set up and solve the innermost integral of Eq.~(\ref{eq:Gani}). We can use the same substitution as described in Eq.~\ref{eq:sigma_eq} except that now we have a factor of $Z$ as defined in Eq.~(\ref{eq:Z}). This brings in extra partial derivatives that we did not have when simplifying $G^{\rm iso}$, so we cannot take the same route as in Section~\ref{sec:int-Giso}. To deal with these partial derivatives, we can substitute the angular momentum $iL_y$ operator for $v_x{\partial}/{\partial v_z}-v_z {\partial}/{\partial v_x}$ in $Z$. This, combined with the symmetry of the operator $n_z + (iD_\theta/kv)\nabla^2_{\rm ang}$, allows us to simplify the inner integral:
\begin{equation}
\int_{S^2} d\Omega~n_x \left( n_z + \frac{iD_\theta}{kv}\nabla^2_{\rm ang}\right)^{-1}Z =
\int_{S^2} d\Omega~\sigma(-iL_y) f_0^{ani} .
\end{equation}
We can use the spherical harmonic decomposition of the phase space density to re-write this,
\begin{equation}
f_0({\boldsymbol v}) = \sum_{\ell=0}^\infty \sum_{m=-\ell}^\ell a_{\ell m}(v) Y_{\ell m}(\hat{\boldsymbol v}).
\end{equation}
(Here $f_0^{\rm ani}$ is the same sum, but without the $\ell=0$ term.)
We further use the relation that $-iL_y = \frac12(L_--L_+)$, where $L_\pm$ are the raising and lowering operators; and the orthogonality rules for the spherical harmonics, $\int d\Omega~ Y_{\ell m} Y_{\ell'm'} = (-1)^m \delta_{\ell,\ell'} \delta_{m,-m'}$. Making these substitutions in the integral, we have
\begin{eqnarray}
\int_{S^2} d\Omega~n_x \left( n_z + \frac{iD_\theta}{kv}\nabla^2_{\rm ang}\right)^{-1}Z &=&
\frac{1}{2} \sum_{\ell=1}^\infty \sum_{m=-\ell}^\ell (-1)^m \sigma_{\ell,-m} \bigl[\sqrt{\ell(\ell+1)-m(m+1)}a_{\ell,m+1}
\nonumber \\ && ~~~~~~~~~~~~
- \sqrt{\ell(\ell+1)-m(m-1)}a_{\ell,m-1}\bigr].
\end{eqnarray}
The anisotropy in $f_0^{\rm ani}$ is a result of photoionization. Non-relativistic photoionization of H and He is an electric dipole process (the electron goes from a bound s orbital to an unbound p wave), and hence the differential cross section has contributions up to $\ell=2$. This means that the only value of $\ell$ that is important to this integral is $\ell=2$. Using Eq.~(\ref{eq:sig-components}), we can write everything in terms of $\sigma_{2,1}$:
\begin{equation}
\int_{S^2} d\Omega~ \sigma(-iL_y) f_0^{ani}  = \sqrt{6}~a_{2,0} \sigma_{2,1} - a_{2,2}\sigma_{2,1}-a_{2,-2}\sigma_{2,1}.
\label{eq: sigma_21}
\end{equation}

Substituting the result from Eq.~(\ref{eq: sigma_21}) into Eq.~(\ref{eq:Gani}), we have
\begin{equation}
G^{\rm ani}(0) = \frac{1}{\bar{n_e}} {\int_0^{\infty}} dv~v^2~ \sigma_{2,1} (\sqrt{6}~a_{2,0} - a_{2,2}-a_{2,-2}).
\label{eq:Gani-sol}
\end{equation}
Now that the integral is only in terms of $v$, we can integrate to find $G^{\rm ani}(0)$ --- but this procedure still requires the anisotropic distribution function $a_{2,m}(v)$ as input.

\subsection{Computing the anisotropic distribution function: methods}
\label{sec: ani_methods}

We model the anisotropic part of the particle distribution with a Fokker-Planck equation:
\begin{equation}
\frac{\partial f}{\partial t} = {\boldsymbol\nabla}_{\boldsymbol v}\cdot \left(
-{\boldsymbol A}f + {\bf D}{\boldsymbol\nabla}_{\boldsymbol v}f
\right) + S,
\end{equation}
where derivatives are in velocity space, ${\boldsymbol A}$ is the mean acceleration, and ${\bf D}$ is the $3\times 3$ diffusion tensor, and $S$ is the source term (the number of electrons produced per unit physical volume, per unit volume in velocity space, per unit time; the units are m$^{-6}$\,s$^{2}$). Writing this in terms of multipole moments, this is
\begin{equation}
\frac{\partial a_{\ell,m}(v)}{\partial t} =
\frac{1}{v^2} \frac{\partial}{\partial v} \left( -A(v)\ v^2\ a_{\ell,m}(v)
+ D_{\parallel\parallel}(v)\ v^2\ \frac{\partial a_{\ell,m}(v)}{\partial v}
\right)
- \ell(\ell+1)D_{\theta}(v) a_{\ell,m}(v)
+ S_{\ell,m}(v).
\label{eq: partial_a_lm}
\end{equation}

We first need the source term $S({\boldsymbol v})$.
The electrons are produced by photoionization, so there is a mapping between initial photon energy and final electron energy:
\begin{equation}
E_\lambda = I_X + \frac12 m_ev^2,
\label{eq:EIX}
\end{equation}
where $X\in\{{\rm H,He}\}$ indicates which element is ionized to get the electron, $I_X$ is its ionization energy, and $\frac12m_ev^2$ is the kinetic energy of the outgoing electron. Here $E_\lambda$ is the energy of a photon in bin $\lambda$.

Following the notation of Ref.~\cite{Zeng_2021}, the number of ionizations per unit volume per unit time per photon energy bin $\Delta E$ is $n_{\rm H}FA_{j\lambda}$, where $F$ is the incident flux of ionizing photons (in photons\,m$^{-2}$\,s$^{-1}$); and ${\cal A}_{j\lambda}$ is the fraction of the ionizing photons that are absorbed in slab $j$ and are in photon energy bin $\lambda$ (this has $\sum_{j\lambda}{\cal A}_{j\lambda}=1$). Then the source term is
\begin{equation}
S = \sum_{X\in\rm\{H,He\}} n_{\rm H} F {\cal A}_{j\lambda} \frac{\tau_{j\lambda}^{X}}{\tau_{j\lambda}^{\rm HI}+\tau_{j\lambda}^{\rm HeI}} \frac1{v^2\Delta v}\frac{d{\rm Prob}}{d\Omega},
\end{equation}
where the ratio of optical depths $\tau$ is the fraction of those absorptions that are by species $X$; $v$ is the velocity bin width $\Delta v = \Delta E/(m_ev)$ corresponding to the photon energy bin width $\Delta E$ (this relation is the derivative of Eq.~\ref{eq:EIX}); and $d{\rm Prob}/d\Omega$ is the angular probability distribution for the emitted electron (in sr$^{-1}$); and $v^2\Delta v\,d\Omega$ is a volume in velocity space (units: m$^3$\,s$^{-3}$).

Photoionization of H or He from a collimated unpolarized source produces an outgoing p wave with angular momenta $m=\pm 1$ (each with 50\% probability and no correlation); thus $d{\rm Prob}/d\Omega$ is $\propto |Y_{11}|^2 + |Y_{1,-1}|^2$, and so the normalized distribution is
\begin{equation}
\frac{d\rm Prob}{d\Omega} =\frac3{8\pi} \sin^2\theta
= \frac3{8\pi} \left[\frac{2}{3} (4\pi)^{1/2}Y_{0,0} - \frac{1}{3} \left(\frac{16\pi}{5}\right)^{1/2}Y_{2,0}\right].
\end{equation}
This can then be substituted back into the source term. Since $F$ is the incident flux of ionizing photons, we know
\begin{equation}
F = \frac{U n_H (1+f_{He})}{1-U/c},
\end{equation}
where $U$ is the ionization front velocity. 
Substituting everything back into our source equation, we find:
\begin{eqnarray}
\sum_{\ell,m} S_{\ell,m} Y_{\ell,m} &=& \frac{3}{8\pi} \sum_{x\epsilon H,He} n_H \frac{U n_H (1+f_{He})}{(1-U/c)} A_{i\lambda} \frac{\tau^x_{j\lambda}}{\tau^{HI}_{j\lambda}+\tau^{HeI}_{j\lambda}} \frac{m_e}{v \Delta E}
\nonumber \\ && \times
\Bigl[\,\frac{2}{3} (4\pi)^{1/2}Y_{0,0} - \frac{1}{3}\Bigl(\frac{16\pi}{5}\Bigr)^{1/2}Y_{2,0}\,\Bigr].
\end{eqnarray}
Only the $\ell=2, m=0$ term contributes to $G^{\rm ani}$, so we focus on it:
\begin{equation}
S_{2,0} = -\frac{1}{8\pi} \left(\frac{16\pi}{5}\right)^{1/2} \sum_{n_H, H, He} n_H \frac{U n_H (1+f_{He})}{(1-U/c)} A_{i\lambda} \frac{\tau^x_{j\lambda}}{\tau^{HI}_{j\lambda}+\tau^{HeI}_{j\lambda}} \frac{m_e}{v \Delta E}.
\label{eq:S_2,0}
\end{equation}

Now that we have solved for $S_{2,0}$, we can continue with our solution of Eq.~(\ref{eq: partial_a_lm}) for $a_{l,m}$. We discretize the equation in both velocity and time so that we can solve for specific values of $a_{2,0}$ using linear algebra. First, we discretize the partial derivative in time using the backward Euler method (which is stable even for stiff systems): $\partial_t Q(t)\approx [Q(t)-Q(t-\Delta t)]/\Delta t$.
We also write it in a format that makes it more suggestive:
\begin{eqnarray}
v^2 S_{2,0}(v,t) + \frac{v^2 a_{2,0}(v,t-\Delta t)}{\Delta t} &=& -\frac{\partial}{\partial v} \left(-A(v,t) v^2 a_{2,0}(v,t) \right)
\nonumber \\ && - \frac{\partial}{\partial v} \left[ D_{\parallel\parallel}(v,t)\ v^2 \left(\frac{\partial}{\partial v} a_{2,0}(v,t)\right)\right]
\nonumber \\ && + 6 D_{\theta}(v,t) v^2 a_{2,0}(v,t) + \frac{1}{\Delta t}v^2 a_{2,0}(v,t).
\label{eq: continuous_S_2,0}
\end{eqnarray}

Next, we consider how the velocities of this distribution are discretized on the right-hand side. We use the integer subscript $i$ to denote which bin value of $v$ we are considering.
The $S_{2,0}(v,t)$, $a_{2,0}(v,t-\Delta t)$, and $D_\theta(v,t)$ terms are simplest: they do not contain any derivatives, so they can simply be written in terms of each value of $v$:
\begin{equation}
v^2 S_{2,0}(v,t) \rightarrow v_i^2 S_{2,0}(v_i,t) 
~~~{\rm and}~~~
6 D_\theta(v,t) v^2 a_{2,0}(v,t) \rightarrow 6 D_\theta(v_i,t) v_i^2 a_{2,0}(v_i,t).
\label{eq: simple_discrete}
\end{equation}

We can consider the middle term, which has two derivatives in terms of $v$. If we consider a discretized derivative operator using half steps (i.e., the finite difference between $v_i$ and $v_{i+1}$ gives a derivative at $v_{i+\frac12}$), we have a term with $v_{i+\frac{1}{2}}$, a term with $v_{i-\frac{1}{2}}$, and a term with $v_i$. Since the $a_{2,0}$ coefficient of this term has two derivatives acting on it, the discretization simplifies to depend on $a_{2,0}$ at integer bins $v_{i-1}$, $v_i$, and $v_{i+1}$:
\begin{equation}
\frac{\partial}{\partial v} \left[ D_{\parallel\parallel}(v)\ v^2 \left(\frac{\partial}{\partial v} a_{2,0}(v)\right)\right] \rightarrow \frac{F(v_{i+\frac{1}{2}})-F(v_{i-\frac{1}{2}})}{v_{i+\frac{1}{2}}-v_{i-\frac{1}{2}}}
\end{equation}
where
\begin{equation}
    F(v_{i + \frac{1}{2}}) = D_{\parallel \parallel}(v_{i + \frac{1}{2}}) v_{i + \frac{1}{2}}^2 \frac{a_{2,0}(v_{i+1}) - a_{2,0}(v_i)}{v_{i+1}-v_i}.
\label{eq: Half-step_D_para}
\end{equation}

Finally, the first term in Eq. (\ref{eq: continuous_S_2,0}) is more complex than all the others because when we discretize it, $a_{2,0}(v_i)$ will not be on whole number intervals; it will instead be on half steps. This means that when we try to convert our continuous equation into a discrete linear algebra equation, this $a_{2,0}(v_i)$ will not fall into intervals that allow it to be defined by a vector or matrix. This is a common problem when solving advection equations, and the simplest solution is to approximate $A(v_{i+\frac12})a_{2,0}(v_{i+\frac12})$ by a value at an integer data point; the ``look to the right'' approximation (using $v_{i+1}$) is stable for advection toward smaller velocities (negative $A$) and the ``look to the left'' approximation (using $v_i$) is stable for advection toward larger velocities (positive $A$). Since the former is the physical case here, we use that method:
\begin{equation}
    \frac{\partial}{\partial v} \left(-A(v) v^2 a_{2,0}(v) \right) \rightarrow \frac{H(v_{i+1})-H(v_i)}{v_{i+1}-v_i}
\end{equation}
where
\begin{equation}
    H(v_{i + 1}) = -A(v_{i + 1}) v_{i + 1}^2 a_{2,0}(v_{i+1}).
\label{eq: Step_A_a}
\end{equation}
We found this approach to be numerically stable.
Gathering all the terms for the discrete solution to $S_{2,0}(v_i)$, we can write an equation in terms of the substitutions that we made in Eq.~(\ref{eq: Step_A_a}) and Eq.~(\ref{eq: Half-step_D_para}):
\begin{eqnarray}
    v_i^2 S_{2,0}(v_i,t)  + \frac{v_i^2 a_{2,0}(v_i,t-\Delta t)}{\Delta t} &=& -\frac{H(v_{i+1},t)-H(v_i,t)}{v_{i+1}-v_i} - \frac{F(v_{i+\frac{1}{2}},t)-F(v_{i-\frac{1}{2}},t)}{v_{i+\frac{1}{2}}-v_{i-\frac{1}{2}}} 
    \nonumber \\ &&
    + 6 D_\theta(v_i,t) v_i^2 a_{2,0}(v_i,t) + \frac{1}{\Delta t}v_i^2 a_{2,0}(v_i,t).
\label{eq: discrete_S_2,0}
\end{eqnarray}

Since each of these terms have a factor of $a_{2,0}(v_{i+1})$, $a_{2,0}(v_{i-1})$, or $a_{2,0}(v_i)$ in them, we can write an $N\times N$ matrix equation of the following form:
\begin{equation}
\left(\begin{array}{c}
v_1^2 S_{2,0}(v_1,t) + v_1^2 a_{2,0}(v_1,t-\Delta t)/\Delta t\\
v_2^2 S_{2,0}(v_2,t  + v_2^2 a_{2,0}(v_2,t-\Delta t)/\Delta t)\\
\vdots\\
v_N^2 S_{2,0}(v_n,t)  + v_N^2 a_{2,0}(v_N,t-\Delta t)/\Delta t
\end{array}\right) = {\bf M}
\left(\begin{array}{c}
a_{2,0}(v_1,t)\\
a_{2,0}(v_2,t)\\
\vdots\\
a_{2,0}(v_N,t)
\end{array}\right)
\label{eq: lin_alg_solution}
\end{equation}
Here ${\bf M}$ is the matrix that contains all the coefficients of $a_{2,0}(v_i)$. Due to how the derivatives of $a_{2,0}(v_i)$ are arranged in Eq.~(\ref{eq: discrete_S_2,0}), ${\bf M}$ will be a tri-diagonal matrix. Its entries are given by the following equations:
\begin{equation}
M_{i,i} = \frac{-v_i^2A(v_i)}{v_{i+1}-v_i}
+ \frac1{v_{i+\frac12}-v_{i-\frac12}} \left[
\frac{v_{i+\frac12}^2D_{\parallel \parallel}(v_{i+\frac12})}{v_{i+1}-v_i}
+ \frac{v_{i-\frac12}^2D_{\parallel \parallel}(v_{i-\frac12})}{v_i-v_{i-1}} \right]
+ 6 D_\theta(v_i) v_i^2 + \frac{v_i^2}{\Delta t}
\label{eq: diagonal_term}
\end{equation}
for the diagonal terms;
\begin{equation}
M_{i,i+1} = \frac{v_{i+1}^2A(v_{i+1})}{v_{i+1}-v_i} - \frac{v_{i+\frac12}^2 D_{\parallel\parallel}(v_{i+\frac{1}{2}})}{(v_{i+\frac{1}{2}}-v_{i-\frac{1}{2}})(v_{i+1}-v_i)}
\label{eq: upper_diagonal}
\end{equation}
for the above-diagonal terms; and
\begin{equation}
M_{i,i-1} = -\frac{v_{i-\frac12}^2 D_{\parallel\parallel}(v_{i-\frac{1}{2}})}{(v_{i+\frac{1}{2}}-v_{i-\frac{1}{2}})(v_{i+1}-v_i)}
\label{eq: lower_diagonal}
\end{equation}
for the below-diagonal terms. We have $M_{ij}=0$ with $|i-j|\ge 2$.

This solution involves looking back at the previous time $t-\Delta t$; however, in our problem with a steady ionization front propagating at velocity $U$, this is equivalent to looking at the next slab in the ionization front model. The time interval between slab widths is $\Delta t = \Delta N_{\rm H}/(n_{\rm H}U)$, where $\Delta N_{\rm H}$ is the hydrogen column density per slab.
\changetext{An initial condition of $a_{2,0}(v,t_{\rm init})=0$ is set on the neutral side of the front (the anisotropy is zero before illumination by ionizing radiation).}

\subsection{Computing the anisotropic distribution function: results}
\label{sec: ani_res}

\begin{figure}[h]
    \centering
    \includegraphics[width=1\linewidth]{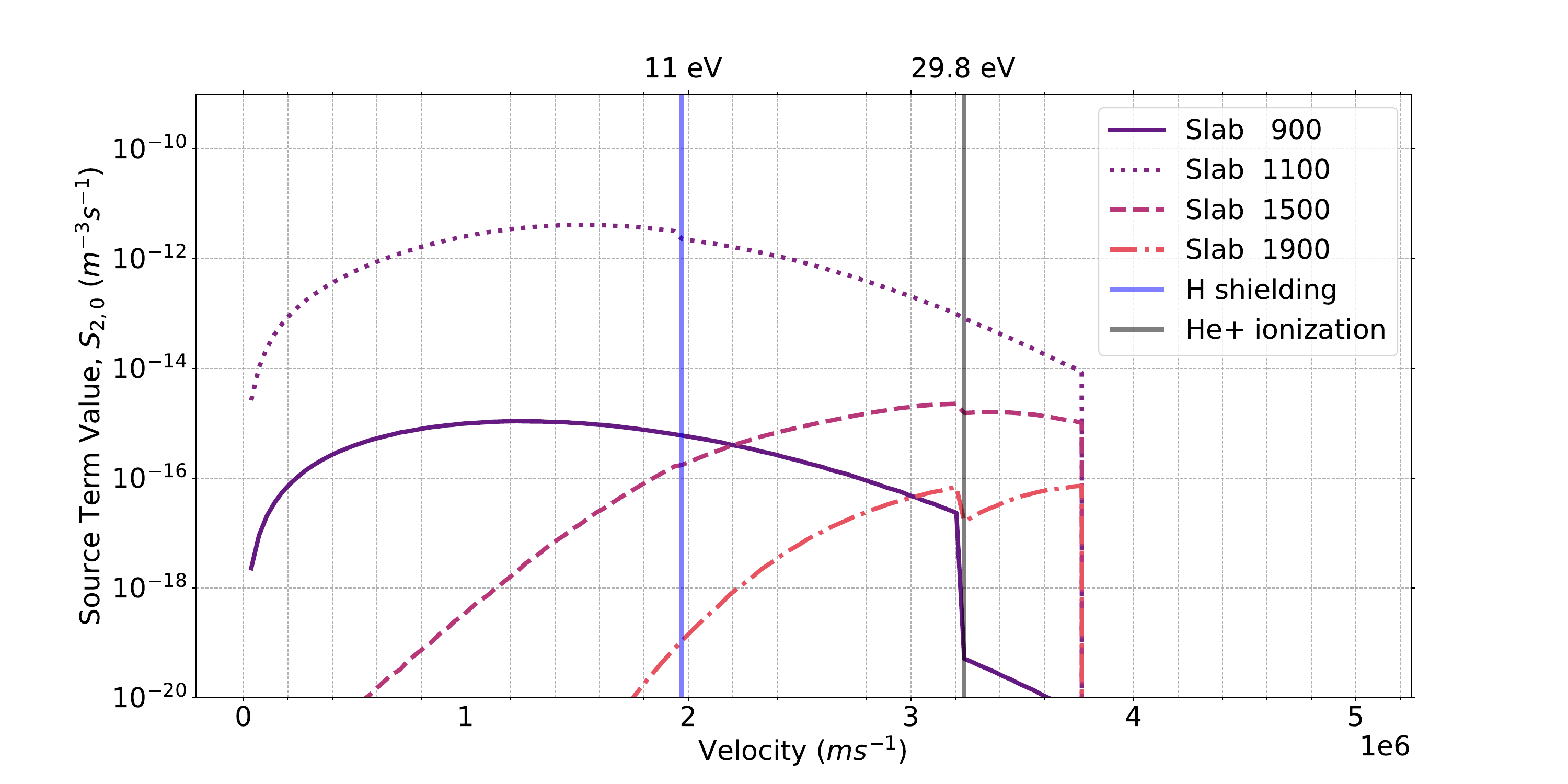}
    \caption{The change in the source term across 142 velocity slabs for the last wavenumber ($k = 10^{-6} ~ \rm{m}^{-1}$) slab (Eq. \ref{eq:S_2,0}). The vertical axis is the value of the source term. The horizontal axis is the velocity of electrons at which we are calculating the source term. Specific reionization front slabs chosen in Figure \ref{fig:im_w_2d} are used to illustrate the change in $S_{2,0}$ through the reionization front. Vertical lines mark the electron velocity at 11.0 eV (the maximum photoelectron velocity from a photon that can ionize H but not He\,{\sc i})  and at 29.8 eV (the maximum photoelectron velocity from a photon that can ionize He\,{\sc i} but not He\,{\sc ii}).}
    \label{fig:S20_plt}
\end{figure}
\begin{figure}[ht!]
    \centering
    \includegraphics[width=1\linewidth]{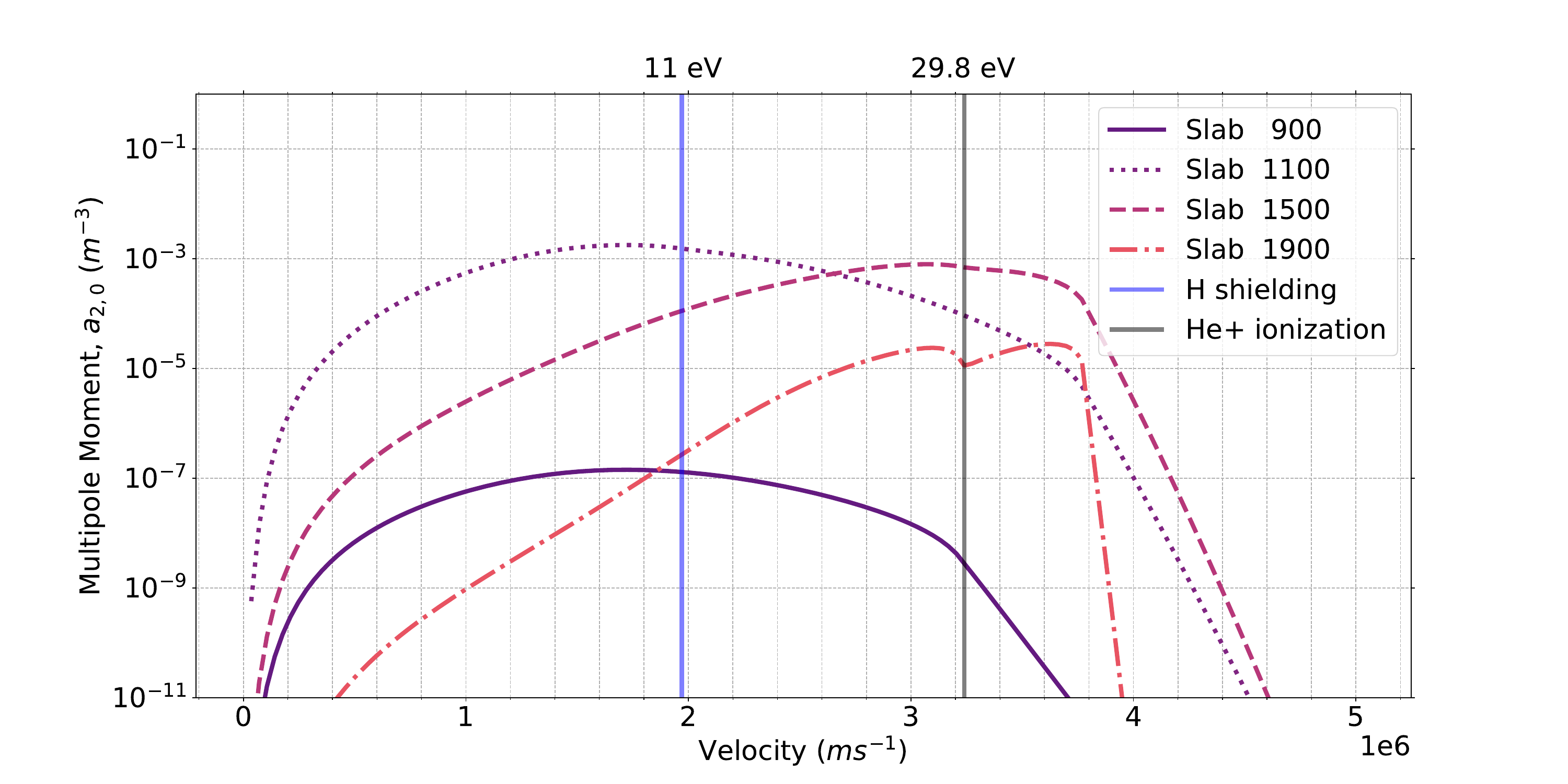}
    \caption{The change in the magnitude of the multipole moment across 142 velocity slabs for the last wavenumber, $k = 10^{-6}\,{\rm m}^{-1}$ (Eq. \ref{eq: partial_a_lm}). The vertical axis is the value of the multipole moment. The horizontal axis is the velocity of electrons at which we are calculating the source term. The change in $a_{2,0}$ is illustrated at the specific reionization front slabs chosen in Figure \ref{fig:im_w_2d}. Vertical lines mark the electron velocity at 11.0 eV (the maximum photoelectron velocity from a photon that can ionize H but not He\,{\sc i})  and at 29.8 eV (the maximum photoelectron velocity from a photon that can ionize He\,{\sc i} but not He\,{\sc ii}). 
    }
    \label{fig:a20_plt}
\end{figure}

While calculating $G^{ani}$, the source term ($S_{2,0}$) and multipole moment ($a_{2,0}$) are the main contributors to the overall behavior. We examine each term's evolution in the front in terms of the velocity slabs and distance slabs. The velocity slabs are discretized values of electron velocity within the front. Each slab is used when calculating the values of each of the terms as the front passes. The distance slabs are discrete values for the distance through the front as given by the reionization front model \cite{Zeng_2021}. In Figures \ref{fig:S20_plt} and \ref{fig:a20_plt}, the slabs plotted and labeled in the legend are those marked in Figure \ref{fig:im_w_2d}. 

{\bfseries{\slshape The source term}}: The source term ($S_{2,0}$) is the flux of electrons (change in electron number density) over a small volume in the reionization front. We examine this term over the distribution of velocities at the same wavenumber ($k = 10^{-6} ~ \rm{m}^{-1}$) and depth in the front. We start from the neutral side (large distances or largest slab number) and work toward the ionized side since this is the direction of evolution experienced by a gas parcel. The source term at 4 such locations is shown in Fig.~\ref{fig:S20_plt}. We note three key velocities where there are features in the source function: $2.0\times 10^6$ m/s (\changetext{kinetic energy 11.0 eV;} the maximum electron velocity from a photon with 13.6--24.6 eV energy, which can ionize H but not He\,{\sc i}); $3.2\times 10^6$ m/s (\changetext{29.8 eV;} the maximum electron velocity from a photon with energy 24.6--54.4 eV ionizing He\,{\sc i}); and $3.8\times 10^6$ m/s (\changetext{40.8 eV;} the maximum electron velocity from a photon with energy $\le 54.4$ eV ionizing H, i.e., the maximum velocity possible from a source that has been shielded by He\,{\sc ii}). Since the angular dependence of the source is the same in all cases, $S_{2,0}$ traces the overall photoelectron injection rate.

Deep in the neutral side (Slab 1900), there is strong shielding so only photons near 54.4 eV make it to this depth; thus the maximum $S_{2,0}$ is around the velocities where these photons produce photoelectrons on He\,{\sc i} ($3.2\times 10^6$ m/s) and H ($3.8\times 10^6$ m/s). As one moves to the left (Slab 1500), the photon spectrum is not quite as hard, and so the step at $3.2\times 10^6$ m/s is less pronounced. As one moves toward the ionized side of the front (Slab 1100), there is less shielding, and due to its larger abundance, H is the main element being ionized. The source term is thus much more broadly distributed between zero and $3.8\times 10^6$ m/s (kinetic energy $54.4-13.6=40.8$\,eV). Even farther into the ionized side (Slab 900), there is a lower source term (since very little neutral material remains), and also a large drop in the electron source term at $3.2\times 10^6$ m/s. This is because of the lower photoionization rate for He\,{\sc i} than H, so much more He\,{\sc i} remains than H, and the maximum velocity of the photoelectrons is $3.2\times 10^6$ m/s (kinetic energy $54.4-24.6=29.8$\,eV).

{\bfseries{\slshape The quadrupole moment in velocity space}}: The multipole moment $a_{2,0}$ is the principal higher-order moment contributing to $G^{\rm ani}$. We examine this term across the velocities in the ionization front to determine how they affect the magnitude of the anisotropic distribution. The change in the multipole moment is illustrated in Figure~\ref{fig:a20_plt} for the largest wavenumber in the simulation, within the free streaming limit ($k = 10^{-6}\,{\rm m}^{-1}$). Each curve tracks the change in the multipole moment at different distances (slabs) into the reionization front.
We examine the evolution of $a_{2,0}$ moving through the reionization front, again starting on the neutral side.

On the neutral side (slabs 1900 and 1500), the maximum of the multipole moment $a_{2,0}$ is at large velocities. Here there are fewer ionizations occurring; they are mostly He ionizations, but there are some H photoionizations, which are responsible for the electrons at velocities of (3.2--3.7)$\times 10^6\,{\rm m}\,{\rm s}^{-1}$. Although these higher-energy electrons slow down and thermalize, they also isotropize as they slow down, leading to smaller quadrupole moments as the velocity decreases. Since these electrons are interacting with a finite-temperature plasma, there is an exponential tail of electrons at {\em higher} energies than the maximum photoionization energy that are accelerated by ``lucky'' collisions before slowing down. On the ionized side of the front (slab numbers 1100 and 900), all energies of photons from the source are streaming through the gas, allowing both H and He to be ionized. The electron distribution, including its quadrupole moment, has a larger distribution of lower velocities from the ionization of H. The velocity where the multipole moment is at maximum is lower than $2.0\times 10^6$ m s$^{-1}$, the velocity above which H photoelectrons are suppressed because He can absorb those photons instead. This is expected due to the larger abundance of H atoms and the large photoionization cross-section at the lower energies.



\begin{figure}[ht!]
    \centering
    \includegraphics[width=1\linewidth]
    {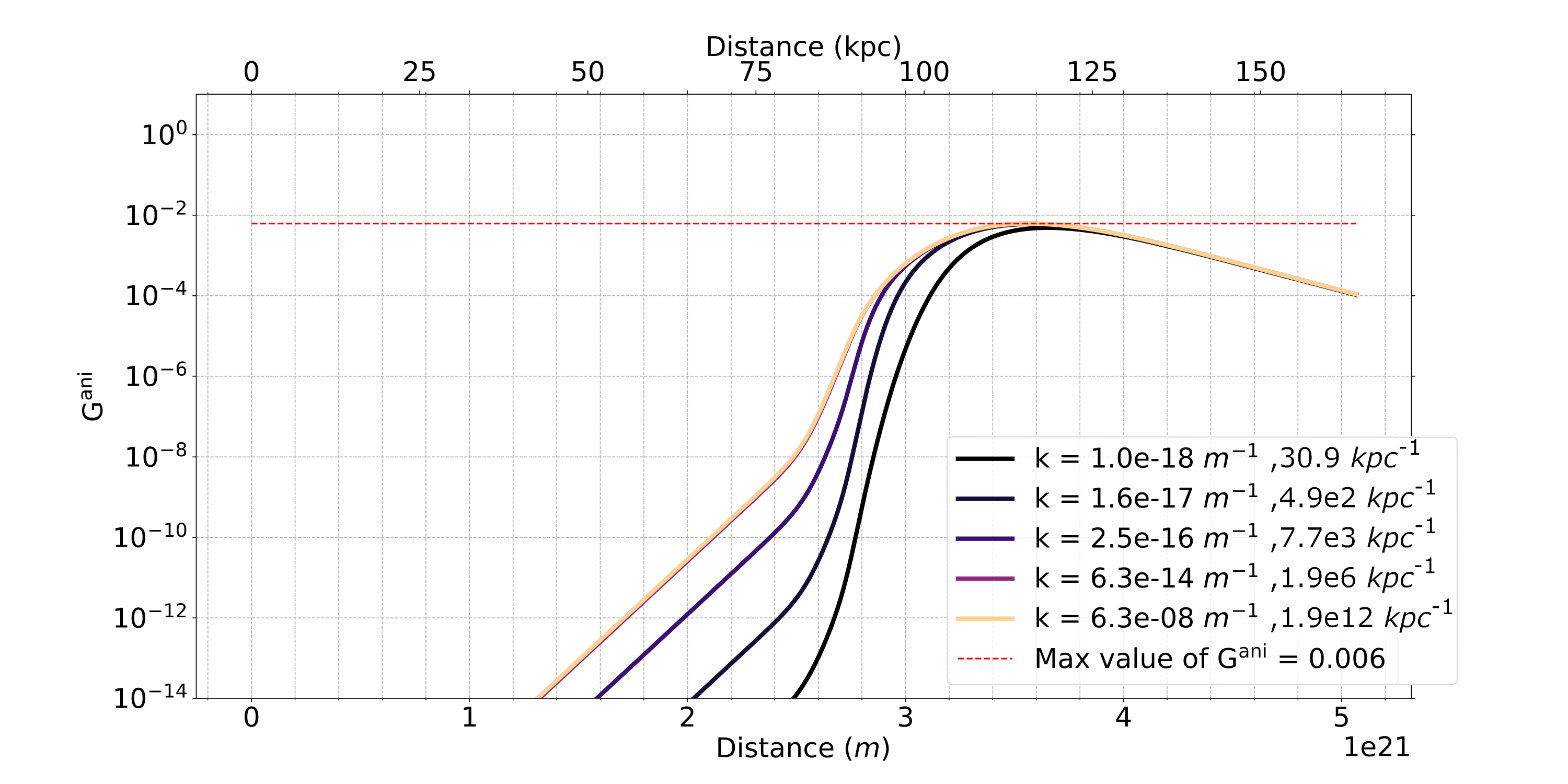}
    \caption{The change in the $G^{\rm ani}$ term across wave number (k) slabs (Eq. \ref{eq:Gani-sol}). The vertical axis is the value of the anisotropic part of the distribution. The horizontal axis is the distance through the reionization front. The increasing wavenumber is inversely proportional to the distance scale it is examining.  We integrated $G^{\rm ani}$ over 142 velocity bins for this simulation.}
    \label{fig:gani_k_slabs}
\end{figure}

{\bfseries{\slshape The anisotropic source for Weibel instability}}: We now compute $G^{\rm ani}$ and plot the output to investigate how the reionization front evolves in Figure~\ref{fig:gani_k_slabs}. The resulting figure has very similar behavior to that of ${G^{\rm iso}(u)}/{u}$ in Figure~\ref{fig:giso_k_slabs} in the H shielding region, but it only reaches non-negligible values for a small portion of the ionization front.

On the neutral side (distance $>4\times 10^{21}\,$m), $G^{\rm ani}$ is small but increasing (toward the left). The anisotropy is dominated by the highest-energy electrons, which are in the free-streaming limit.
Moving toward the ionized side (or equivalently forward in time), $G^{\rm ani}$ reaches maximum at a value close to order $1 \times 10^{-3}$ at distance $\approx 3.5\times 10^{21}\,$m, where the helium ionization fraction is still $\sim 1\%$ and the hydrogen ionization fraction has risen to $\sim 0.1\%$. This rapid growth is a strong indicator that the perturbation will be capable of generating a magnetic field in the H shielding region. This will be further analyzed in Section~\ref{sec: results}. 

Finally, close to the ionization source, $G^{\rm ani}$ is very small for all wavenumbers due to the gas being largely ionized (hence shorter mean free path) and having a very small source term (since it is photoionization itself that produces an anisotropic electron velocity distribution). Smaller wavenumbers exhibit a similar staggered behavior to those in the $G^{\rm iso}$ plot, while the larger wavenumbers can mostly be represented by one line. As explained in Section \ref{sec:int-Giso}, these differences are due to the smaller wavenumbers being in the collisional limit (perturbation wavelength $\gg$ mean free path) and the large wavenumbers being in the free-streaming limit (perturbation wavelength $\ll$ mean free path). However, unlike in $G^{\rm iso}$, both the small and large wavenumbers follow roughly the same pattern. The large differences seen with $G^{\rm iso}$ only appear in $G^{\rm ani}$ close to the ionization source.

\section{The instability growth rate}
\label{sec: results}

As mentioned above in Section \ref{sec:int-Gani}, we examine the rapid growth of the anisotropy and evaluate if it is fast enough to generate a magnetic field. The rate of growth of the magnetic field is given by the imaginary part of $\omega$, \changetext{which we obtain by substituting $u=\omega/k$ into Eq.~(\ref{eq:Gk-eq}) and solving for $\omega$:}
%
\begin{equation}
\Im \omega = k  \Im u
 = \frac{k}{\Im [G^{\rm iso}(u)/u]} \left( G^{\rm ani}(0) - \frac{k^2}{k_{\rm sd}^2} \right),
\label{eq: Im_w_growth_rate}
\end{equation}
\changetext{where we recall that $G^{\rm ani}(0)$ is real but $G^{\rm iso}(u)/u$ is purely imaginary.}
At the smallest scales, the magnetic field will decay the fastest. If we consider the case where $G^{\rm ani} \rightarrow 0$, the growth rate will be dominated by $k^3$. The wavenumber, $k$, is the inverse of the length scale so this results in the instability on the smallest scales decaying quickly. The peak growth rate will move up the length scale over the duration of the reionization front.

\begin{figure}[ht!]
    \centering
    \includegraphics[width=1\linewidth]{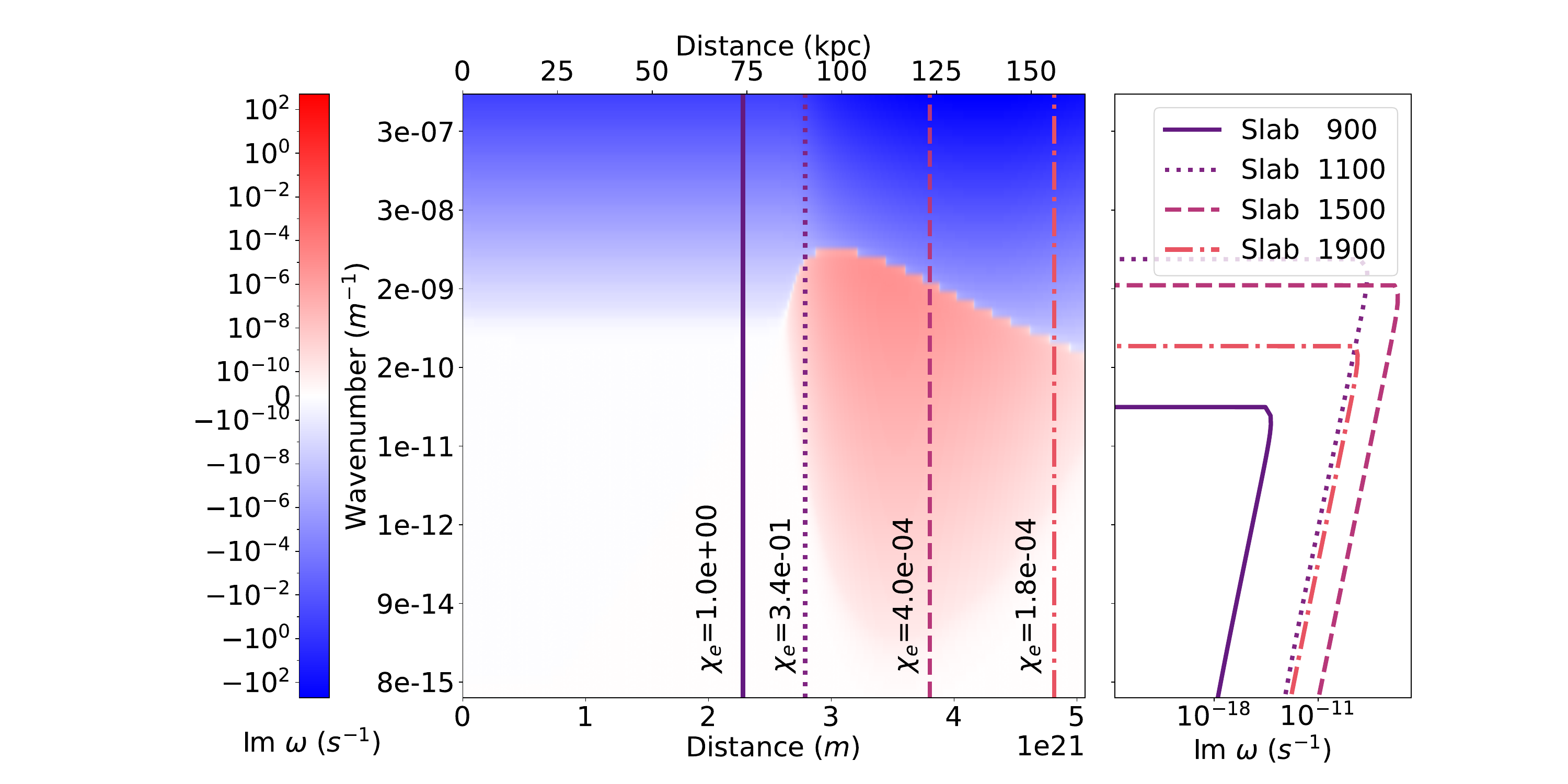}
    \caption{This plot illustrates the change in $\Im \omega$ (Eq. \ref{eq: Im_w_growth_rate}), which measures the speed at which the anisotropy grows. This is measured across wave number (k) slabs and distances through the reionzation front. The main panel illustrates the value of $\Im \omega$ across the ionization front and wavenumbers. The vertical lines in this plot show the distances at which we will examine the behavior of the anisotropic growth rate. They are labeled by the fraction of free electrons at that distance. The right panel takes slices at distances marked in the main plot to illustrate the change in the anisotropic growth.}
    \label{fig:im_w_2d}
\end{figure}

In Figure~\ref{fig:im_w_2d}, we show a 2D plot of the $\Im \omega$ values over all wavenumbers and slabs in the ionization front in the simulation. This plot quantifies the rate of the growth of the anisotropy at different wavenumbers (distance scales) and different distances through the reionization front. As we move through the front, starting at the source, all growth rates are zero or negative up to where the H shielding starts. Past this point, the magnitude of the growth rates quickly increases. 

Throughout the entire front, all wavenumbers greater than about $3\times10^{-8} ~ \rm{m}^{-1}$ stay negative. At these wavenumbers, the distance scale is of a small enough magnitude to not have anisotropies. However, examining larger distance scales, the rapid growth of the anisotropy is clear. The largest growth occurs a short distance from the beginning of the H shielding region. The speed of this growth is about $6 \times 10^{-6} ~ \rm{s}^{-1}$ and approximately occurs at a wavenumber of $3 \times 10^{-9} ~ \rm{m}^{-1}$ with a distance of about $3\times10^{21}$ m. After this point, the growth rates gradually decrease for larger wavenumbers and turn negative at larger distances through the front.

At the point where the growth of the anisotropy peaks, it has a time scale of about $1.7 \times 10^5 ~ $s. The reionization front passage time is approximately $2 \times 10^{14} ~ $s, many orders of magnitude longer than the time it takes for the anisotropy to grow at $k \approx 3 \times 10^{-9} ~ {\rm m}^{-1}$. Though we assume a linear growth scale for this analysis, the conclusion from these results is clear. Magnetic fields generated within this front will have sufficient time to grow and establish before the front moves on. After the front, the magnetic fields on the smaller scales where the growth rate was highest will decay rapidly. However, the fields generated on the larger scales will experience little to no decay, so they will be locked into the background for long timescales.

\begin{figure}[ht!]
    \centering
    \includegraphics[width=1\linewidth]{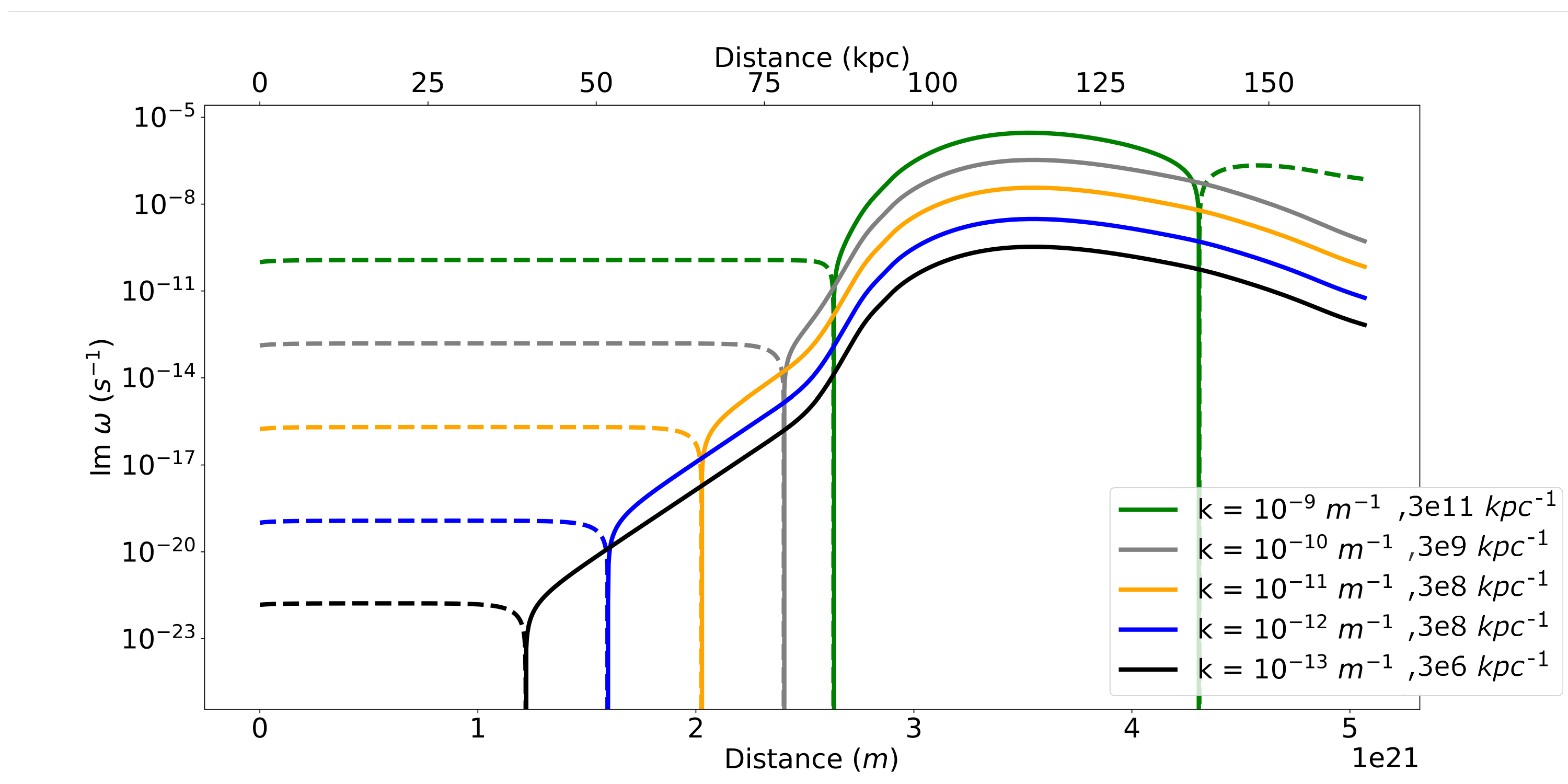}
    \caption{This plot shows the change in $\Im \omega$ over distance slabs for a few wavenumbers. The dashed lines illustrate decay in the anisotropy, while the solid lines show growth for the same wavenumber. After passage of the front, all scales are decaying, but fields at scales $k\lesssim 5\times 10^{-12}\,{\rm m}^{-1}$ will survive for a cosmologically long time.}
    \label{fig:im_w_distance}
\end{figure}

Decay and growth timescales are illustrated in Figure \ref{fig:im_w_distance}. Dashed lines are decay, while solid lines show growth of the anisotropy and magnetic field for wavenumbers near the maximum growth timescale. For wavenumbers ($k = 10^{-9} ~ {\rm m}^{-1}$) which produce values close to the maximum of $\Im \omega$, the timescale for peak growth is roughly equivalent to the decay timescale. Any magnetic fields produced at this length scale will grow and decay quickly, making them poor candidates for seed fields.

Moving to smaller wavenumbers --- longer length scales --- the magnitude of the growth timescale decreases slightly while the maximum decay timescale decreases quickly. For the smallest wavenumber we examined in this plot ($k=10^{-13} ~ \rm{m}^{-1}$), the maximum growth rate ($\approx 10^{-10} ~ \rm{s}^{-1}$) is more than 10 orders of magnitude greater than the largest decay timescale, $\sim 10^{-22} ~ {\rm s}^{-1}$. For this length scale, the anisotropy will grow quicker than the reionization front passage time, while its decay will take longer than the current age of the universe. This is true for the two lowest wavenumbers shown in the figure, making them excellent candidates for the length scales at which long-lived seed magnetic fields could form.

The wavenumbers \changetext{$k\gtrsim 10^{-11}\,{\rm m}$ or length scales $k^{-1}\lesssim 1\,$AU have decay timescales that} are shorter than the age of the universe. \changetext{These} largest wavenumber here will not \changetext{by themselves} produce a long-lasting magnetic field. 
We emphasize that the above rates are in {\em linear} theory: given the short growth timescales, the Weibel instability will inevitably enter the non-linear regime. Detailed investigation of this regime is critical, but left to future work. \changetext{Other non-linear phenomena would also be essential to transferring this power to larger scales, as suggested by the gamma-ray constraints.\footnote{\changetext{A magnetic field with a very short coherence length needs to be stronger in order to achieve the same net deflection angle, so lower limits on the intergalactic field typically scale as $B_{\rm min}\sim \lambda_B^{-1/2}$, where $\lambda_B$ is the coherence length \cite[e.g.][]{2010Sci...328...73N}.}} Our seeding mechanism naturally produces large-scale ``headless vector'' alignment, since the linear instability produces ${\boldsymbol B}$ parallel to the ionization front; but the gamma ray constraints require the directionality (e.g., $+x$ vs. $-x$) to be coherent over some scale $\lambda_B$. This would require a dynamo mechanism (e.g., from turbulence \cite{Cain2025}), but we defer to future work a detailed investigation of whether the Weibel instability seeding combined with the hydrodynamical turbulence (also from reionization) can produce a field structure consistent with all of the observational constraints.}

\section{Discussion}
\label{sec: discussion}

Our results suggest that the epoch of reionization --- a period that unfolded after the emergence of the first stars and galaxies, but before the assembly of large-scale structure --- may naturally 
host conditions capable of generating cosmological magnetic seed fields. Reionization is often described in terms of steadily expanding ionized bubbles around early galaxies, but the actual morphology of these ionization fronts is highly irregular. They are neither planar nor smooth; instead, they exhibit strong spatial variations and dynamical instabilities driven by inhomogeneous density, velocity fields, and anisotropic illumination.

Motivated by this complexity, we explored whether such front-level instabilities can act as sites of magnetic field generation. In particular, the geometry of the radiation field is 
asymmetric, with ionizing photons predominantly arriving from one direction. Combined with the quadrupolar angular dependence of the photoionization cross-section, this induces an anisotropic 
distribution of electron velocities within the front. Under these circumstances, the Weibel instability can operate efficiently, converting velocity-space anisotropies into magnetic fields.

Using a simulated ionization front, we quantified both the isotropic and anisotropic components of the electron velocity distribution. The anisotropy becomes especially pronounced in the central region of the front, reaching values as high as $G^{\rm ani}(0)\sim 6\times 10^{-3}$. Such large departures from isotropy are sufficient to drive rapid Weibel growth. Indeed, the linear growth timescale we obtain is extremely short --- $2\times 10^5~\mathrm{s}$, which is orders of magnitude smaller than the front's characteristic crossing time of roughly $\sim {\rm few} \times 10^{14}\,{\rm s}$ --- indicating that magnetic fields can be generated essentially 
``on the fly'' as the front propagates. 

These findings imply that reionization fronts may contribute to the early seeding of cosmic magnetism, \changetext{if these seed fields can survive and be} amplified by other processes like the turbulent dynamo \cite{zhou2023magnetogenesiscollisionlessplasmaweibel, Cain2025}. For instance, \cite{Cain2025} demonstrated that ionization fronts generated during cosmic reionization strongly heated the surrounding intergalactic medium (IGM). This rapid heating left the gas in a state of pronounced pressure imbalance. As the system attempted to re-establish pressure equilibrium, the resulting motions likely induced small-scale turbulent flows (even in regions below mean density \cite{2018MNRAS.474.2173H, Cain2025}), which could then 
act to amplify pre-existing intergalactic magnetic fields. Conversely \cite{Liu_2025} examines the inverse transfer process and finds that the amplification of the seed field may be suppressed. However, they work with small-scale field strengths that are strong enough to tie electrons to field lines, $kr_{\rm g}<1$ (where $r_{\rm g}$ is the gyroradius), whereas in our problem the electrons are only weakly perturbed in traversing one wavelength of our magnetic perturbations until the field strength rises to $\sim 10^{-10}\,$G (if that happens at all), so the phenomenology may be very different.\footnote{In the notation of \cite{Liu_2025}, for a thermal electron, $kr_{\rm g} = \sqrt{3T_0/2\sigma}\,kd_0$; in our case, for a fastest-growing wavenumber of $k\sim 10^{-9}\,{\rm m}^{-1}$ and electron velocities $v_e\sim 3\times 10^6\,$m/s, we have $kr_{\rm g}\sim 2\times 10^{-10}\,$G$/B$.} 

In our analysis, we have focused on the linear phase of the Weibel instability, where the electron anisotropy grows without significant magnetic back-reaction. Once the instability amplifies, the magnetic field reaches a characteristic strength; however, the field becomes strong enough to smear out the electron velocity distribution, signaling the onset of the non-linear regime. At this stage, deflection by the stochastic magnetic field begins to isotropize the electrons more efficiently than Coulomb collisions.
We discuss some aspects of the onset of non-linearity in Appendix~\ref{ss:nonlin}, but a quantitative determination of the saturation amplitude and post-saturation behavior requires 
going beyond the linear theory.

This non-linear evolution is critical to the ultimate question of whether magnetic fields generated by the Weibel instability can survive for cosmological timescales. The optimistic possibility is that as magnetic fields grow and contribute to the effective angular diffusion rate (i.e., there is a $D_{\theta,B}$, the electron quadrupole anisotropy decreases but never vanishes (it declines only as $\propto D_{\theta,B}^{-1}$ since the source $S_{2,0}$ is governed by photoionization physics and cannot be shut off). This would cause small-scale modes to decay, but allow larger-scale modes to continue to grow (albeit more slowly). The result would be that --- even deep in the ionization front where the gas is still mostly neutral --- the magnetic field is weak but the peak of its power moves toward large scales. If perturbations can move from scales of $k^{-1}\sim 3\times 10^8\,{\rm m}$ where the growth rate peaks to $k^{-1} > 10^{11}\, {\rm m}$, then with the smaller skin depth post-reionization, these fields could survive for a cosmologically long time.

The next step is to develop a quasi-linear model that captures the feedback between the growing magnetic field and the electron velocity distribution. Such an approach is needed because the 
full non-linear evolution operates over a broad hierarchy of spatial and temporal scales: field fluctuations at scales of a few $\times 10^8\,{\rm m}$ and growth timescales of $\sim 10^5\,{\rm s}$ must be modeled, but the front passage takes a few $\times 10^{14}\,{\rm s}$ during which time an electron moves by a few $\times 10^{20}\,{\rm m}$.
Direct simulations that simultaneously resolve the microphysical Weibel growth and the global front structure are therefore prohibitively 
expensive with current computational capabilities. A quasi-linear or hybrid kinetic treatment offers a tractable path forward, allowing the key elements of magnetic back-reaction and 
electron isotropization to be incorporated without requiring an unrealistically large dynamic 
range.

Additional processes may also become important as the system approaches saturation. Scattering by any cosmic rays produced in the post-front environment, and the 
hydrodynamic relaxation that follows the passage of the ionization front \cite{2018MNRAS.474.2173H, Cain2025} can all contribute to electron isotropization. These effects may either suppress or further enhance the Weibel-generated 
fields, depending on the local plasma conditions. Determining when such processes become dominant—and how they modify the magnetic-field evolution—will require a dedicated study of the non-linear development of the instability, which we defer to future work.

\section*{Data availability}

The code used to generate the results in this paper is located on GitHub at \url{https://github.com/Jorie286/Reionization-Magnetic-Fields}.

\acknowledgments

We thank Molly Wolfson for useful discussions and suggestions.

During the preparation of this work, the authors received support from the Simons Foundation (award 60052667) and NASA (award 22-ROMAN11-0011).
M.R. acknowledges support from the CCAPP fellowship at The Ohio State University.
J.M. received summer research support from the Captain Forrest R. Biard Undergraduate Research Scholarship Fund in Physics.

This work was supported in part by an allocation of computing time from the Ohio Supercomputer Center's Cardinal cluster \citep{Ohio_Supercomputer_Center2024-dl}.

\appendix

\section{Velocity diffusion tensor}
\label{sec:vd}

In the main text, we needed the velocity diffusion tensor to assess the 2-body relaxation of the electron distribution function. The angular diffusion coefficient is related to the transverse component of the velocity tensor by a factor of $v^2$ (the conversion from transverse velocity to angular deflection). The general formula for species $a$ in a background of isotropic thermal plasma components is
\begin{equation}
D_{\theta,a}(v) =
\frac{D_{a,\perp\perp}(v)}{v^2} = \sum_b \frac{q_a^2q_b^2 n_b \ln\Lambda_{\rm C}}{8\pi \epsilon_0^2m_a^2 v^3} \left[
\left(1-\frac{\sigma_b^2}{v^2}\right)\,{\rm erf}\,\frac{v}{\sqrt2\,\sigma_b} + \sqrt{\frac 2\pi} \frac{\sigma_b}{v} e^{-v^2/2\sigma_b^2}
\right],
\label{eq:ang-diff}
\end{equation}
where the sum over $b$ is over species; $\ln\Lambda_{\rm C}$ is the Coulomb logarithm; and $\sigma_b = \sqrt{k_{\rm B}T_b/m_b}$ is the velocity dispersion \citep[\S1.5.3]{1983hppv.book.....G}.\footnote{Beware of the factor of 2 difference between our definition of the diffusion coefficient and Ref.~\cite{1983hppv.book.....G}, which includes a factor of $\frac12$ in the diffusion term of the Boltzmann equation.} Note that the quantity in brackets is $\propto v$ at small $v$ (the $1/v$ terms cancel) so the angular diffusion coefficient is $\propto 1/v^2$. The quantity in brackets goes to 1 at large velocities. The Coulomb logarithm in a singly ionized plasma with $k_{\rm B}T\lesssim\,$Ry (so collisions are in the classical regime) can be expressed in convenient units as:
\begin{equation}
\ln\Lambda_{\rm C} = 
\frac32 \ln \frac{k_{\rm B}T}{\rm Ry} - \frac12 \ln (64\pi a_0^3n_e),
\end{equation}
where Ry is the Rydberg energy and $a_0$ is the Bohr radius.

The mean acceleration of a fast electron (drag force divided by electron mass) is given by the Chandrasekhar dynamical friction formula,
\begin{equation}
A_a(v) = - \sum_b \frac{q_a^2q_b^2n_b(m_a/m_b+1) \ln\Lambda_{\rm C}}{4\pi\epsilon_0^2m_a^2v^2}\left[
{\rm erf}\,\frac{v}{\sqrt2\,\sigma_b}
- \sqrt{\frac 2\pi}\frac{v}{\sigma_b} e^{-v^2/2\sigma_b^2}
\right],
\end{equation}
and the along-path part of the diffusion coefficient is
\begin{equation}
D_{a,\parallel\parallel}(v) = \sum_b \frac{q_a^2q_b^2n_b(m_a/m_b+1)\sigma_b^2 \ln\Lambda_{\rm C}}{4\pi\epsilon_0^2m_am_bv^3}\left[
{\rm erf}\,\frac{v}{\sqrt2\,\sigma_b}
- \sqrt{\frac 2\pi}\frac{v}{\sigma_b} e^{-v^2/2\sigma_b^2}
\right]
\end{equation}
(in accordance with the Einstein relations that the contributions from species $b$ to $D_{a,\parallel\parallel}(v)$ and $A_a(v)$ differ by a factor of $-k_{\rm B}T_b/m_av$).

\section{Limiting cases of the anisotropy function $\sigma$}
\label{sec:calc_sigma}

In Section~\ref{sec: evaluating_Gs} we discuss the isotropic and anisotropic components of $G$, which led us to an equation form that has one term in common, $\sigma$ (see Eq.~\ref{eq:sigma_eq}). Here we explore the behavior of $\sigma$ in two limiting cases: the free-streaming and collisional cases.

In the free streaming limit where the angular diffusion of electrons is much less than their velocity, we consider the case where $D_\theta/kv \ll 1$. To conveniently write $\sigma$ in this limit, we define the following
\begin{equation}
    \epsilon = -\frac{D_\theta}{k v} \nabla^2_{\rm ang}.
\label{eq: epsilon_free_streaming}
\end{equation}
Note that $\nabla^2_{\rm ang}$ has negative eigenvalues, so for a small diffusion coefficient, $\epsilon$ can be thought of as having an infinitesimal positive value.
This allows us to easily consider small $\epsilon$ when thinking about the behavior in this limit. So, we get the following equation for $\sigma$ in the free streaming limit:
\begin{equation}
    \sigma \rightarrow \left(n_z - i \epsilon\right)^{-1} ~ n_x, ~~~ \frac{D_\theta}{kv}\ll 1.
\label{eq: sigma_free_streaming}
\end{equation}

In the collisional limit where the angular diffusion of electrons is much larger than their velocity, so $D_\theta/kv \gg 1$. In this limit, the $n_z$ term is negligible because $D_\theta/kv\gg 1$ is large; we then recall that $n_x$ is an eigenfunction of $\nabla_{\rm ang}^2$ with eigenvalue $-2$. So, we find
\begin{equation}
    \sigma \rightarrow \frac{i k v n_x}{2 D_\theta}, ~~~ \frac{D_\theta}{kv}\gg 1.
\label{eq: sigma_collisional}
\end{equation}

\subsection{Isotropic contribution: $\sigma_{1,1}$}
\label{sec:sigma_iso}

In the free streaming limit, we use Eq.~(\ref{eq: sigma_free_streaming}): and make the substitution of $n_z = \mu=\cos\phi$ and $n_x = \sqrt{1-\mu^2} \cos\phi$, so $d\Omega = d\mu\,d\varphi$. Then we find
\begin{equation}
\sigma_{1,1} \rightarrow \int Y_{1,1}^\ast\,(n_z-i\epsilon)^{-1} n_x\,d\Omega
= \int_0^{2\pi} \int_{-1}^1 
-\sqrt{\frac{3}{8\pi}} \frac{1-\mu^2}{\mu- i\epsilon} e^{i\varphi} \cos \varphi \,
d\mu \,d\varphi.
\end{equation}
The $\varphi$ integral integrates to $\int_0^{2\pi} e^{i\varphi}\cos\varphi\,d\varphi = \pi$, and we have:
\begin{equation}
\sigma_{1,1} \rightarrow -\sqrt{\frac{3\pi}8} \left[ \ln (\mu-i\epsilon) - \frac12\mu^2 \right]_{-1}^1
\rightarrow -\sqrt{\frac{3\pi}8} \pi i,
\end{equation}
where we take the branch of the natural logarithm corresponding to the path of integration.

\begin{figure}[ht!]
    \centering
    \includegraphics[width=0.95\linewidth]{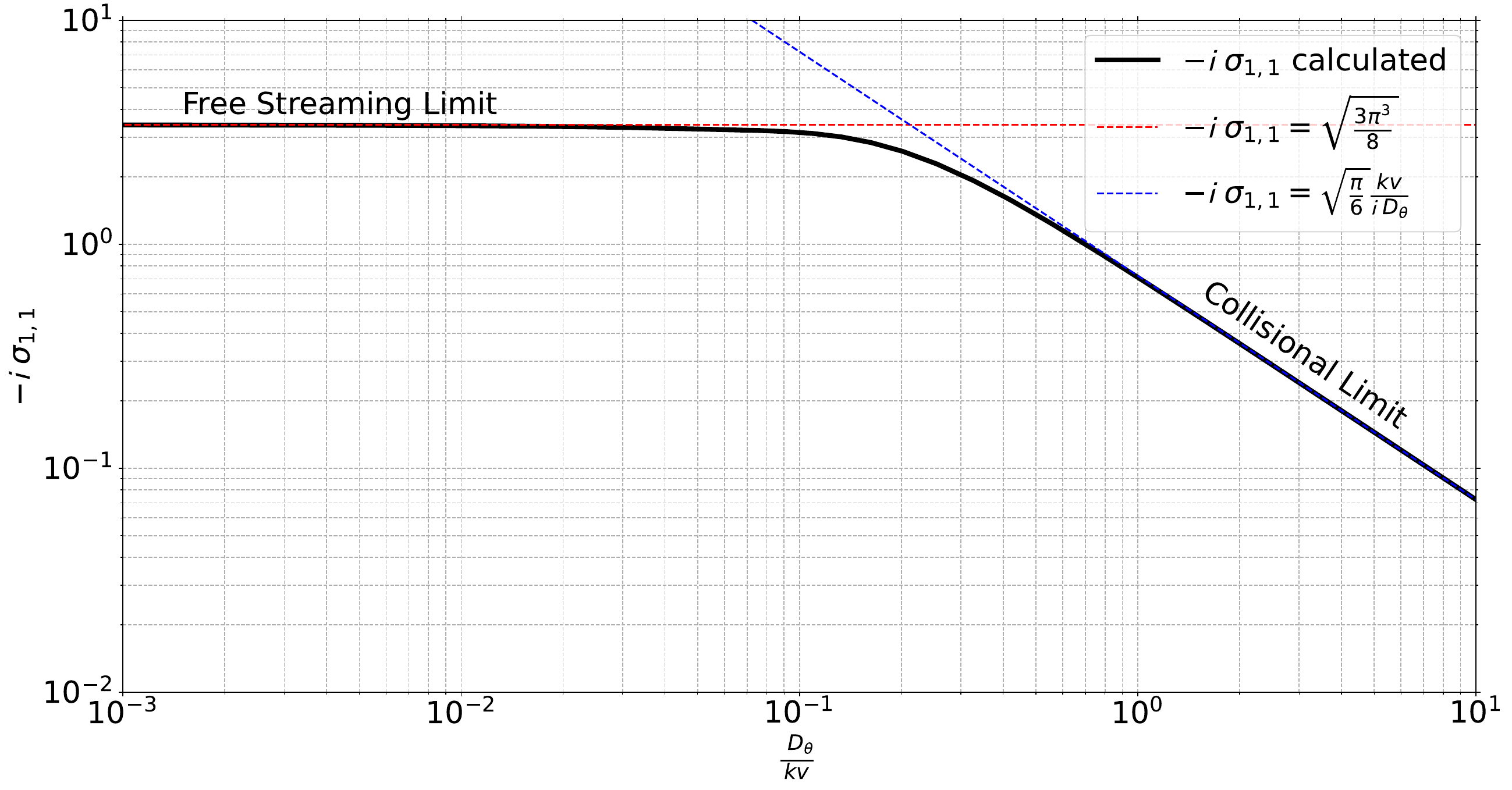}
    \includegraphics[width=0.95\linewidth]{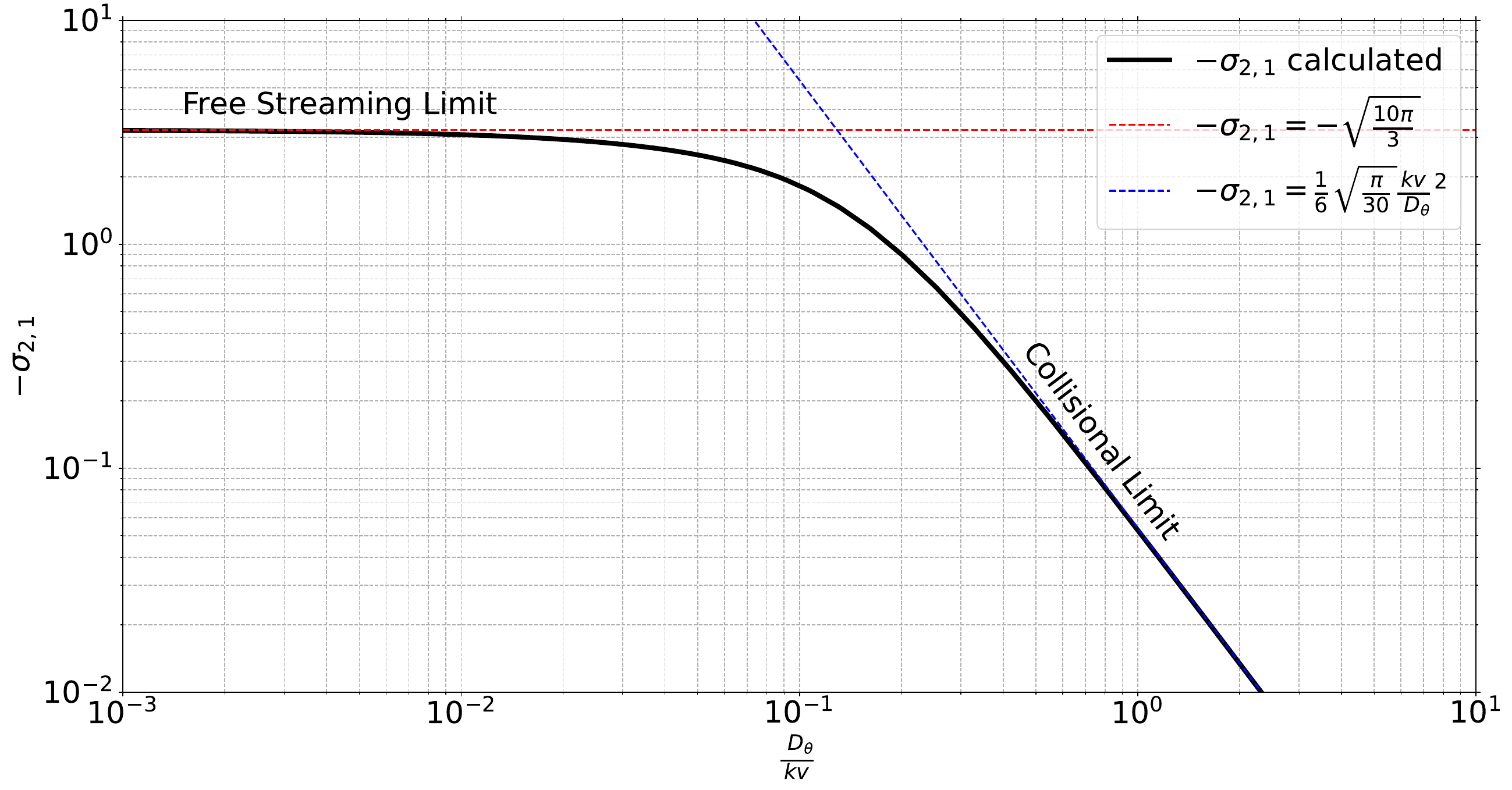}
    \caption{Change in the isotropic contribution ($\sigma_{1,1}$, Eq. \ref{eq: sigma_11}; top panel) and the anisotropic contribution ($\sigma_{2,1}$, Eq. \ref{eq: sigma_21}; bottom panel) as a function of $D_\theta/kv$. Free streaming and collisional limits are marked to illustrate long-term behavior.}
    \label{fig:sigma_11_plt}
\end{figure}

In the collisional limit, we substitute this decomposition into the equation for $\sigma$ (Eq.~\ref{eq: sigma_collisional}); we find that the upper limit for $\sigma_{1,1}$ is
\begin{equation}
    \sigma_{1,1} \rightarrow - \sqrt{\frac{\pi}{6}}\frac{i k v}{D_\theta}.
\end{equation}
These limiting cases are illustrated in the top panel of Figure~\ref{fig:sigma_11_plt}.

\subsection{Anisotropic contribution: $\sigma_{2,1}$}
\label{sec:sigma_ani}

We now wish to find $\sigma_{2,1}$ in both limiting cases. It is possible to do this by integral methods as done for $\sigma_{1,1}$, however, in the collisional case this requires going to higher order than Eq.~(\ref{eq: sigma_collisional}). A simpler alternative is to use Eq.~(\ref{eq: sigmas_sol}) with the case of $\ell=1$ and use the value of $\sigma_{1,1}$ in each limit to solve for $\sigma_{2,1}$. The $\sigma_{0,1}$ will not contribute to this solution (it formally is multiplied by zero), so we find
\begin{equation}
    \sigma_{2,1} = - \sqrt{\frac{10 \pi}{3}}
    ~~({\rm free-streaming})~~~{\rm and}~~~
    \sigma_{2,1} = -\left(\frac{kv}{D_\theta}\right)^2 \frac{1}{6}\sqrt{\frac{\pi}{30}}
    ~~({\rm collisional}).
\end{equation}

These behaviors are shown in the bottom panel of Figure~\ref{fig:sigma_11_plt}. Similar to the behavior of $\sigma_{1,1}$ shown above, $\sigma_{2,1}$ is constant in the free streaming limit and a decreasing function in the collisional limit. However, in the collisional limit, $\sigma_{2,1}$ has a much heavier dependence on $\frac{i~D_\theta}{kv}$, which will result in a rapid decrease in $G^{\rm ani}$.

\section{Non-linear magnetic field growth considerations}
\label{ss:nonlin}

In the calculations of the main text, we assume that the anisotropy will grow linearly. However, this assumption eventually becomes incorrect due to the rapid growth rate demonstrated in Section~\ref{sec: results}. The Weibel instability generally becomes non-linear when the magnetic field itself smears out the initial anisotropy in the electron distribution. Thus we estimate the angular diffusion rate and use this to determine how rapidly isotropy will be induced on the magnetic field. The angular diffusion coefficient is of order
\begin{equation}
    D_\theta \sim \frac{\Delta \theta^2}{t},
\end{equation}
where $\Delta \theta \sim \omega_{\rm c} t$ is a small deflection angle when the electron passes through a magnetic field structure. The cyclotron frequency is $\omega_{\rm c} = \frac{q B}{m_e}$. We can therefore make the approximation that the magnetic contribution to the diffusion coefficient is
\begin{equation}
    D_{\theta,B} \approx \left(\frac{q B}{m_e}\right)^2 \frac{\ell}{v},
\end{equation}
where $\ell$ is the typical size of the structures.

When the magnetic field becomes strong enough that the magnetic contribution to $D_\theta$ exceeds that from Coulomb collisions, linear growth around the background described in the main text will no longer fully describe the system. We can estimate the magnitude of the magnetic field at this critical point --- we will call this $\sigma_{\rm B,crit}$ --- using the angular diffusion equation. Rearranging some terms, we find
\begin{equation}
    \sigma_{\rm B,crit} \sim \frac{m_e}{q} \sqrt{\frac{k_{\rm sd} v D_\theta}{k_{\rm sd} \ell}}.
\end{equation}

Formatting $\sigma_{\rm B,crit}$ as shown above allows us to make some convenient simplifying assumptions. The angular diffusion can be approximated as $\Gamma_2 \approx 10^{-11} \,{\rm s}^{-1}$ and $k_{\rm sd} v \approx (q/m_e) \sqrt{\mu_0 n_e T_e}$ using the definition of $k_{\rm sd}$. The value of  $k_{\rm sd} v$ at $\chi_e = 10^{-2}$ is about $2.3 \times 10^{-3}\, {\rm s}^{-1}$. We scale our results to reference values of $\chi_e \approx10^{-2}$ and the value of $k_{\rm sd} \ell \approx 1$. The magnitude of the magnetic field can thus be approximated as
\begin{equation}
    \sigma_{\rm B,crit} \approx 10^{-14} {\rm G} ~ \left(k_{\rm sd} \ell\right)^{-1/2} \left(\frac{\chi_e}{10^{-2}}\right)^{3/4}.
\label{eq: nonlin_mag}
\end{equation}

The magnetic field generated by the Weibel instability eventually reaches a larger magnitude than the critical magnitude calculated above. This demonstrates that the rapidly growing magnetic field created by the Weibel instability will require further examination in the non-linear regime. 

\bibliographystyle{JHEP}
\bibliography{main.bib}

\end{document}